\documentclass[twoside]{book}
\usepackage{ml-books-english} 

\usepackage{amsmath}
\usepackage{mathtools}
\usepackage{bm}
\usepackage{booktabs}
\usepackage{siunitx}
\usepackage{enumitem}

\usepackage[utf8]{inputenc}

\usepackage[table]{xcolor}

\usepackage[authoryear,round]{natbib}

\usepackage{url}

\usepackage[colorlinks,
linkcolor={red!70!black},
citecolor={blue!60!black},
urlcolor={blue!80!black}]{hyperref}

\newcommand{\avg}[1]{\left\langle#1\right\rangle} 
 
\newcommand{\chapterauthors}[1]{%
  \par\vspace{1em}
  {\Large\bfseries
  \begin{minipage}{\textwidth}
    \raggedright
    #1
  \end{minipage}\par}
  \vspace{1em}
}

\usepackage{wrapfig}

\newtcolorbox{mauricionotebox}[1][Note]{%
    title=Box~\themauricionote. #1,
    colback=white,
    colframe=lightgray,
    colbacktitle=lightgray,
    coltitle=black,
    boxrule=0.5pt,
    arc=0pt,
    left=2mm,
    right=2mm,
    top=1mm,
    bottom=1mm,
    before skip=8pt,
    after skip=8pt,
    fonttitle=\bfseries
}

\newcounter{mauricionote}
\newenvironment{mauriciofloatingnote}[2][tb]
{%
\refstepcounter{mauricionote}%
\begin{boxedfigure}[#1]
\phantomcaption
\begin{mauricionotebox}[#2]
}
{%
\end{mauricionotebox}
\end{boxedfigure}
}

\usepackage{silence}
\usepackage{tikz}

\newcommand{\plotKernelRect}{%
  \begin{tikzpicture}[baseline=-0.5ex, scale=0.4]
    \draw[->, gray, thin] (-0.5, 0) -- (2.5, 0);
    \draw[->, gray, thin] (0, -0.2) -- (0, 1.5);
    \draw[thick, blue] (-0.5, 0) -- (0, 0) -- (0, 1) -- (1.5, 1) -- (1.5, 0) -- (2.3, 0);
  \end{tikzpicture}%
}

\newcommand{\plotKernelExp}{%
  \begin{tikzpicture}[baseline=-0.5ex, scale=0.4]
    \draw[->, gray, thin] (-0.5, 0) -- (2.5, 0);
    \draw[->, gray, thin] (0, -0.2) -- (0, 1.5);
    \draw[thick, red, domain=0:2.2, samples=30] plot (\x, {1.2*exp(-\x)});
    \draw[thick, red] (-0.5, 0) -- (0, 0);
  \end{tikzpicture}%
}

\newcommand{\plotKernelAlpha}{%
  \begin{tikzpicture}[baseline=-0.5ex, scale=0.4]
    \draw[->, gray, thin] (-0.5, 0) -- (2.5, 0);
    \draw[->, gray, thin] (0, -0.2) -- (0, 1.5);
    \draw[thick, red, domain=0:2.2, samples=30] plot (\x, {8*\x*exp(-2.5*\x)});
    \draw[thick, red] (-0.5, 0) -- (0, 0);
  \end{tikzpicture}%
}

\newcommand{\plotKernelGauss}{%
  \begin{tikzpicture}[baseline=-0.5ex, scale=0.4]
    \draw[->, gray, thin] (-1.5, 0) -- (1.5, 0);
    \draw[->, gray, thin] (0, -0.2) -- (0, 1.5);
    \draw[thick, green!60!black, domain=-1.4:1.4, samples=30] plot (\x, {1.2*exp(-\x*\x/0.3)});
  \end{tikzpicture}%
}

\newcommand{\plotGLPhiFunc}{%
  \begin{tikzpicture}[baseline=-0.5ex, scale=0.4]
    \draw[->, gray, thin] (-0.5, 0) -- (4.5, 0);
    \draw[->, gray, thin] (0, -0.2) -- (0, 1.5);
    
    \node[below, font=\scriptsize] at (4.5, 0) {$V$};
    \node[left, font=\scriptsize] at (0, 1.7) {$\Phi$};
    \node[below, font=\scriptsize] at (1, 0) {$\theta$};
    \node[below, font=\scriptsize] at (3, 0) {$\theta+1/\Gamma$};
    \node[left, font=\scriptsize] at (0, 1) {$1$};
    
    \draw[gray, dashed] (3, 0) -- (3, 1);
    \draw[gray, dashed] (0, 1) -- (4.3, 1);
    
    \draw[thick, purple] (-0.5, 0) -- (1, 0) -- (3, 1) -- (4.3, 1);
  \end{tikzpicture}%
}

\newcommand{\VE}{V^{\rm E}}
\newcommand{\VI}{V^{\rm I}}
\newcommand{\XE}{X^{\rm E}}
\newcommand{\XI}{X^{\rm I}}
\newcommand{\NE}{N^{\rm E}}
\newcommand{\NI}{N^{\rm I}}
\newcommand{\rhoE}{\rho_{\rm E}}
\newcommand{\rhoI}{\rho_{\rm I}}
\newcommand{\PhiE}{\Phi_{\rm E}}
\newcommand{\PhiI}{\Phi_{\rm I}}
\newcommand{\WEE}{G^{\rm EE}}
\newcommand{\WEI}{G^{\rm EI}}
\newcommand{\WIE}{G^{\rm IE}}
\newcommand{\WII}{G^{\rm II}}
\newcommand{\VbE}{V^{b,{\rm E}}}
\newcommand{\VbI}{V^{b,{\rm I}}}
 
\newcommand{\Iext}{I^{\mathrm{ext}}} 

\NewDocumentCommand{\IextE}{o}{%
  I^{\mathrm{ext}}_{\mathrm{E}\IfValueTF{#1}{, #1}{}}%
}
\NewDocumentCommand{\IextI}{o}{%
  I^{\mathrm{ext}}_{\mathrm{I}\IfValueTF{#1}{, #1}{}}%
}

\begin{document}

\contribauth{Carvalho et al.}

\chapter{Mean-field theory of neural networks: from spikes to firing rates}
\label{ch:firingrates}

\chapterauthors{
Tawan T. A. Carvalho,\textsuperscript{1}\\
Matheus Haas,\textsuperscript{1}\\
Denis D. Caprioti,\textsuperscript{1}\\
Luis A. R. Delgado,\textsuperscript{1}\\
Artyom Kuznetsov,\textsuperscript{1,2}\\
Henrique L. Schweitzer,\textsuperscript{1}\\
Constantino F. Vasconcelos,\textsuperscript{1}\\
Rodrigo P. Rocha,\textsuperscript{1}\\
Mauricio Girardi-Schappo\textsuperscript{1,*}\\
}

\contribnote{
\textsuperscript{1} NeuroPhysics Lab, Departamento de Física, Universidade Federal de Santa Catarina, Florianópolis SC, Brazil 88040-900\\
\textsuperscript{2} Ural State Medical University, Yekaterinburg, Russia\\
\textsuperscript{*}\href{mailto:m.girardi.schappo@ufsc.br}{m.girardi.schappo@ufsc.br}}

\section*{Abstract}
Connecting microscopic neuronal spiking to macroscopic brain function is a central challenge
in computational neuroscience. Firing-rate models provide a framework for addressing this
problem by describing neuronal activity at an intermediate level between individual spikes
and large-scale population dynamics. Here, we introduce the foundations of firing-rate descriptions,
formalizing their definitions through temporal, population, and trial averaging.
We discuss experimental evidence for firing rates, from the input-output transfer
functions of individual neurons to the tuning curves of different neuronal populations.
We then establish the connection between spikes and firing rates in single neurons
and recurrently connected excitatory-inhibitory populations using mean-field approximation.
We illustrate how collective dynamics emerge from neuronal interactions,
including population stimulus-response curves and bistability induced by recurrent coupling,
and brain rhythms arising from synaptic depression.
We briefly review extensions and applications of firing-rate models to learning,
seizure-like oscillatory activity, visual selectivity, phase-amplitude coupling in neural oscillations,
and whole-brain modeling.
Together, these examples illustrate how firing-rate models provide
effective descriptions of neural activity,
linking the dynamics of individual neurons to collective computations across scales.

\section{Introduction}

\begin{figure*}[t!]
    \includegraphics[width=1\linewidth]{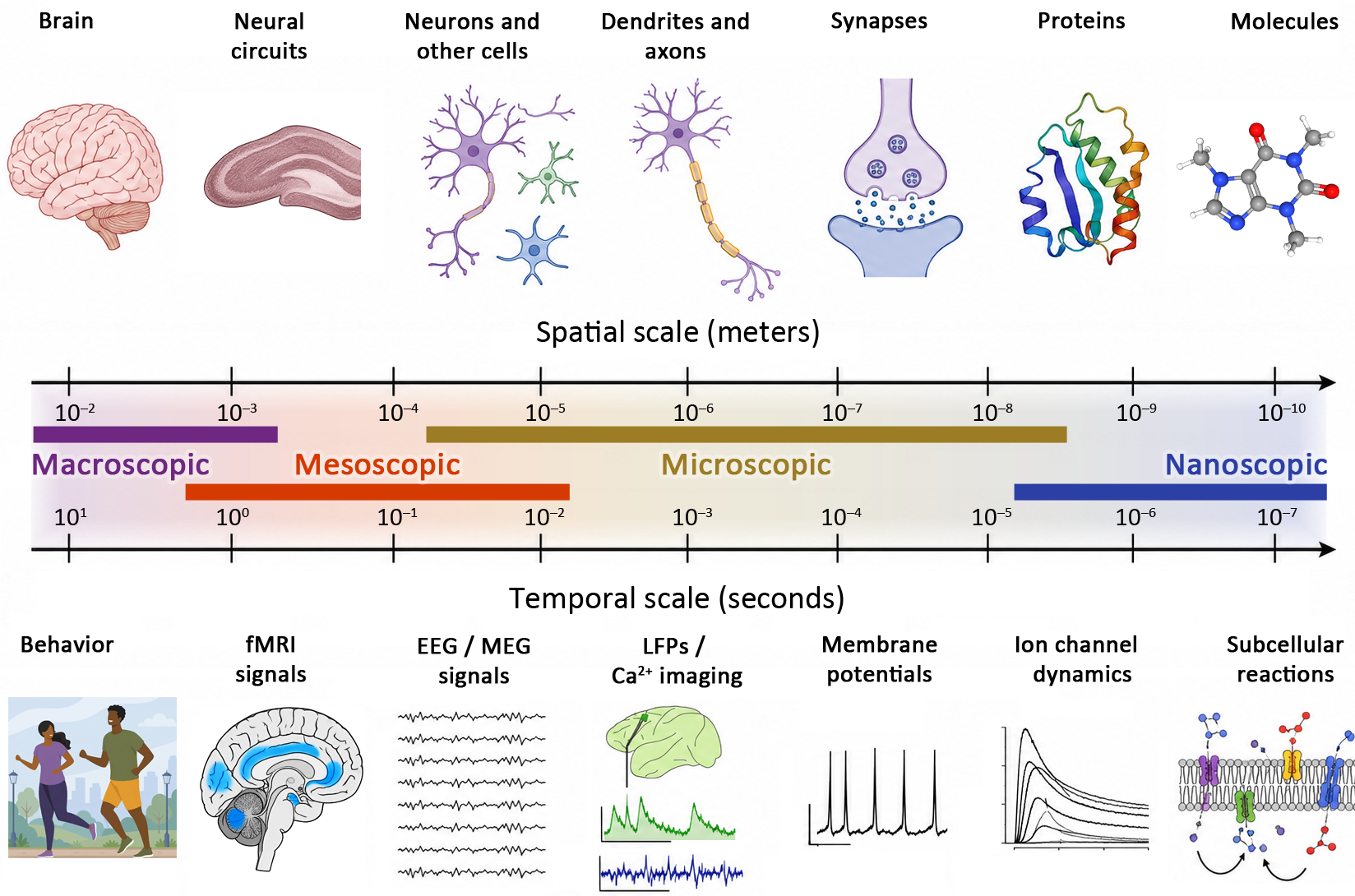}
    \caption{\label{fig:brainscales}Illustration of spatial scales of main components and processes involved
    with or resulting from computations in the brain and their related temporal scales.}
\end{figure*}

The brain operates across an extraordinary range of spatial and temporal scales (Fig.~\ref{fig:brainscales}).
At the largest spatial scales, we can consider the brain as a whole,
or examine the activity of specific regions and neural circuits,
such as the hippocampus (CA1, CA3, and the dentate gyrus), the visual system,
and the cerebellum. Moving toward finer scales, we encounter individual neurons and other brain cells,
including glial cells and astrocytes, followed by their dendrites, axons, and synapses.
At still smaller scales, neural function involves synaptic vesicles, neurotransmitters,
proteins, and ultimately the molecular interactions that sustain cellular activity.

A similarly broad range of timescales characterizes brain dynamics. Behavior unfolds over seconds to hours,
whereas neural activity can be measured at much shorter intervals. Functional magnetic resonance imaging (fMRI),
for example, typically probes brain activity over hundreds of milliseconds to seconds. Electroencephalography (EEG),
magnetoencephalography (MEG), and local field potentials (LFPs) capture electrical
activity on millisecond timescales, while membrane potentials evolve over comparable intervals.
At the cellular and molecular levels, ion channel kinetics and biochemical reactions can occur over microseconds or less.

These scales are not independent. Molecular processes shape the electrical properties of neurons;
neuronal activity, in turn, determines how circuits process and transmit information;
and the collective dynamics of neural circuits contribute to perception, cognition, and behavior.
The relationship also works in the opposite direction: experience and behavior
can induce lasting changes in neural circuits, cellular properties, and even gene expression.
Understanding how these processes interact across scales is therefore a central challenge
in neuroscience~\citep{BlossHunt2019Neurocart}. Different experimental and theoretical approaches
address different parts of this hierarchy. Molecular and cellular studies investigate
the mechanisms underlying neuronal function, while large-scale approaches seek to relate
brain activity to behavior and cognition. Connecting these perspectives requires descriptions
that can bridge the gap between individual neurons and the collective activity of neural populations.

This is the perspective we adopt here. We focus on the intermediate scales connecting
the electrical dynamics of individual neurons to the collective activity of neuronal
populations (see Box~\ref{box:thermo}). Our central interest is in understanding
how the spiking activity of individual neurons gives rise to population-level
dynamics, and how these collective patterns support the processing and representation of information.

\begin{mauriciofloatingnote}[t!]{A thermodynamic analogy}
    \label{box:thermo}
    \begin{wrapfigure}{l}{0.52\linewidth}
    \vspace{-1.5\baselineskip}
    \centering
    \includegraphics[width=\linewidth]{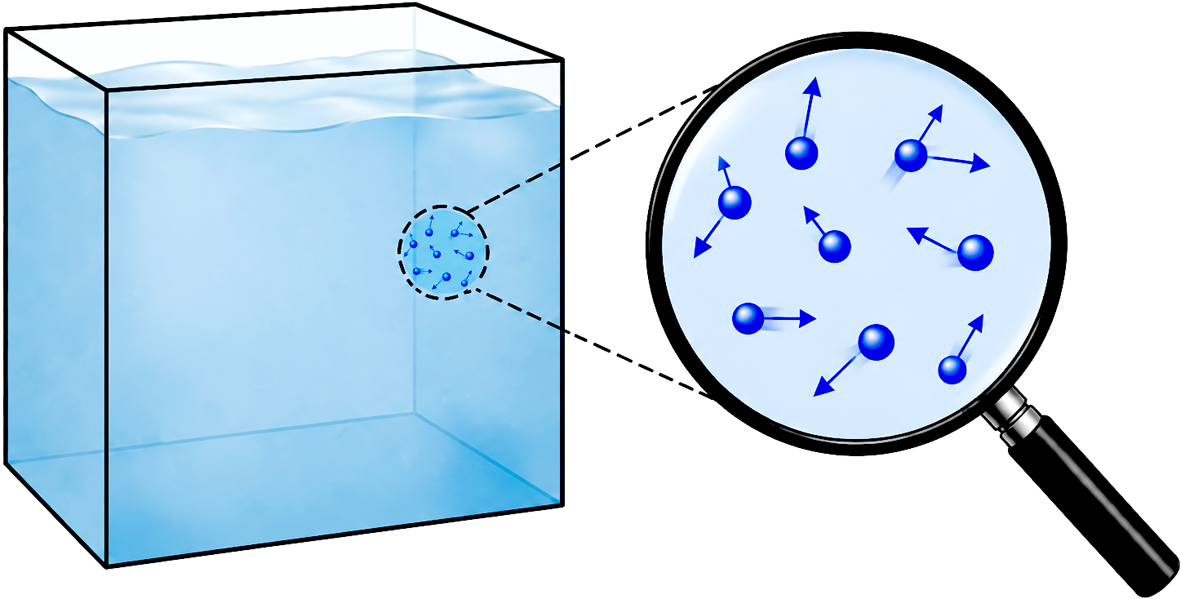}
    \vspace{-\baselineskip}
    \end{wrapfigure}
    An instructive analogy can be made with thermodynamics.
    Macroscopic variables such as temperature, pressure, and volume provide an effective description of a gas,
    even though none of these quantities exist at the level of an individual molecule.
    Instead, we can employ statistical physics techniques to show that they
    emerge from the collective behavior of a very large number of microscopic variables.
    Likewise, spikes may be regarded as the microscopic constituents of neural activity,
    whereas firing rates play the role of macroscopic variables.
    Rather than tracking every individual spike,
    firing-rate models describe the collective behavior emerging from large neuronal populations.
    This analogy also illustrates an important modeling principle.
    Microscopic and macroscopic descriptions are complementary rather than competing.
    The appropriate level of description depends on the question being addressed.
    If the objective is to understand the dynamics of individual neurons,
    spike-based models are more appropriate.
    If the interest lies in the collective activity of large populations,
    firing-rate models often provide a simpler and more useful description.
    Throughout this chapter, we adopt this latter perspective and investigate
    the conditions under which firing rates provide an effective description of neural dynamics.
\end{mauriciofloatingnote}

Neuronal models provide different levels of description for investigating this connection.
They range from detailed representations of neuronal morphology, including dendrites,
the soma, and the axon, to simplified integrate-and-fire (IF) models~\citep{herzNeuron2006,izhikevichDyn,deSchutterBook,Shimoura2021Conn}.
Since we are interested in spikes as fundamental units of
neural information processing~\citep{BialekSpike1996},
we will focus on models that describe electrical activity
through the evolution of the membrane potential,
often represented using equivalent electrical circuits.

The simplest IF model consists of a resistor-capacitor circuit that integrates
incoming inputs until the membrane potential reaches a threshold. A spike is then generated,
and the potential is reset, initiating a new integration cycle. Although highly simplified,
this basic description can be extended to incorporate additional nonlinearities
and adaptation mechanisms, allowing models to reproduce the firing behavior
of hundreds of different neuronal types while remaining computationally
efficient~\citep{Teeter2018GLIF,Trinh2023hMC}. These models provide a
useful starting point for investigating how the intrinsic properties
of individual neurons shape their responses to inputs and, ultimately,
the collective dynamics of neuronal populations.

A natural way to describe this collective activity is through the firing rate:
the average number of spikes generated by a neuron or a population of neurons
per unit time. This description shifts the focus from the precise timing of
individual spikes to the statistical activity of populations, providing
a connection between microscopic neuronal dynamics and larger-scale brain
activity. In particular, we will examine how membrane time constants,
conductances, and adaptive currents influence population firing rates,
and how these collective descriptions can, in turn, provide insight
into the microscopic mechanisms underlying neural computation.

This population-level perspective is also relevant to experimental neuroscience.
Techniques such as fMRI, EEG, MEG, and LFP recordings do not generally resolve
individual neuronal spikes. Instead, they measure signals associated with the
activity of large neuronal populations, reflecting different aspects of their
underlying electrical and metabolic processes. The precise relationship between
these signals and neuronal spiking remains an active area of research.
Nevertheless, their dependence on the collective activity of neurons
provides a natural motivation for describing neural populations
in terms of average firing rates~\citep{Gerstner1992UnivFiringRate}.

\section{What are firing rates?}

\begin{figure*}[t!]
    \includegraphics[width=1\linewidth]{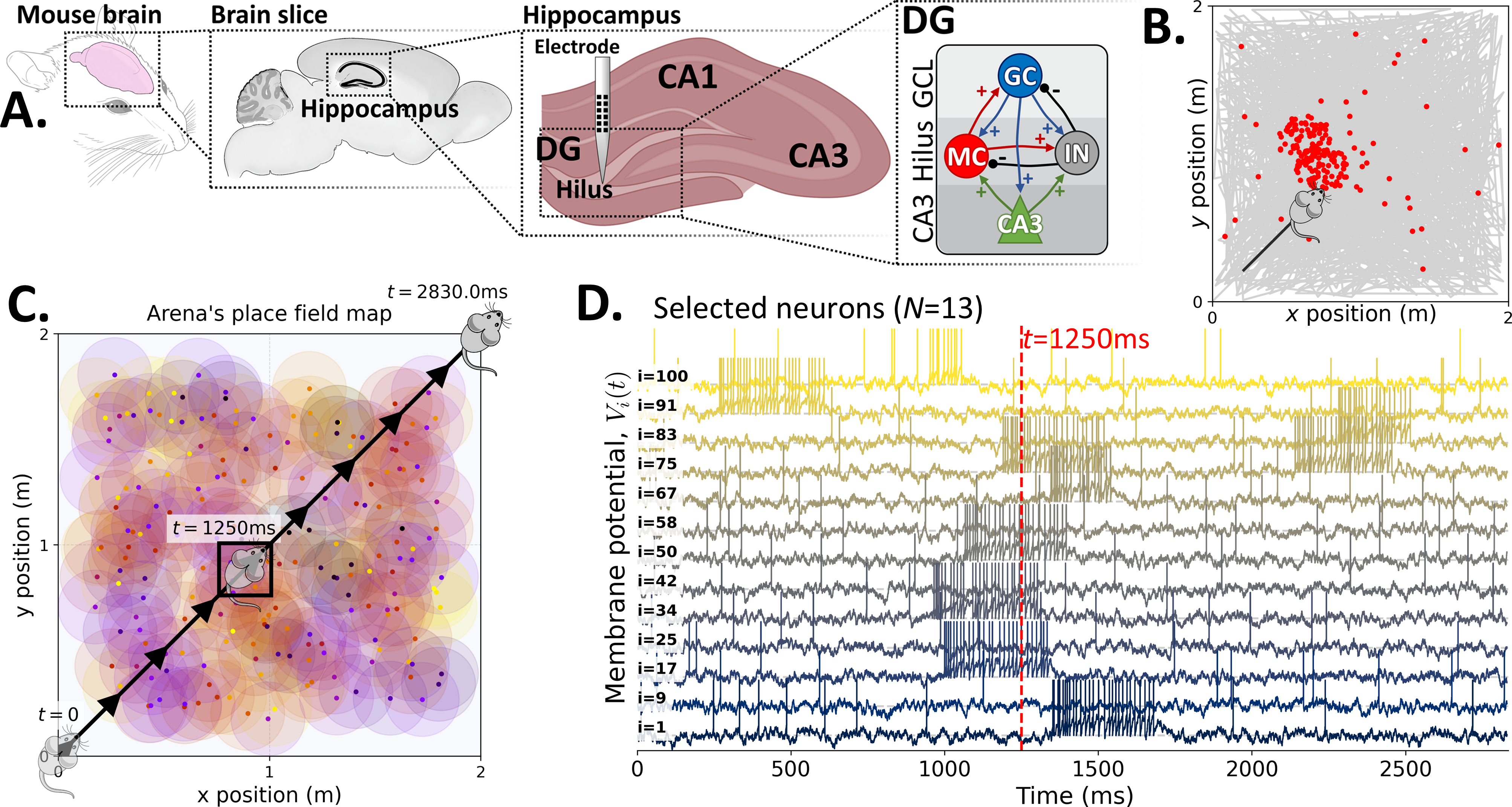}
    \caption{\label{fig:dgplacefields}
Illustration of hippocampal place fields and their relation to firing-rate coding.
\textbf{A.}~Multi-scale representation of the mouse hippocampus, from the whole brain to the dentate gyrus (DG) circuitry,
            highlighting granule cells (GC), mossy cells (MC), interneurons (IN), and CA3 pyramidal neurons~\citep{Hainmueller2020}.
\textbf{B.}~Illustration of spike locations (red dots; trajectory in gray) in a typical recording from a DG place cell during
            a random exploration task~\citep{GoodSmith2017}.
\textbf{C.}~Simulated trajectory of a mouse traversing an arena diagonally.
            Each colored circle represents the place field of a different neuron.
            We selected the instant $t=1,250$~ms to further explore the relation
            between spikes and firing rates (vertical dashed line in panel D, and in Fig.~\ref{fig:frdef}).
\textbf{D.}~Membrane potential traces of selected neurons during the simulation.
            Neurons transiently increase their firing activity when the animal enters their corresponding place fields,
            illustrating how spatial position can be represented by changes in firing rates.
            The simulation is designed for illustration purposes only,
            it is not meant to represent all the biological details of the DG.}
\end{figure*}

One of the motivations for firing-rate models comes from the study of place fields.
Consider a mouse freely exploring an arena while the activity of individual neurons is recorded with an electrode implanted
in the Dentate Gyrus (DG) hilus of the hippocampus (Fig.~\ref{fig:dgplacefields}A).
By correlating the animal's position with neuronal spiking activity,
we observe that some neurons fire preferentially when the animal occupies specific regions of the environment (Fig.~\ref{fig:dgplacefields}B).
These neurons are called \textit{place cells} and their corresponding high-activation zones in space are called \textit{place fields}.
Granule cells (GCs), mossy cells (MCs), CA3 pyramidal neurons are examples of place cells~\citep{GoodSmith2017,Hainmueller2020}.

\begin{figure*}[t!]
    \includegraphics[width=1\linewidth]{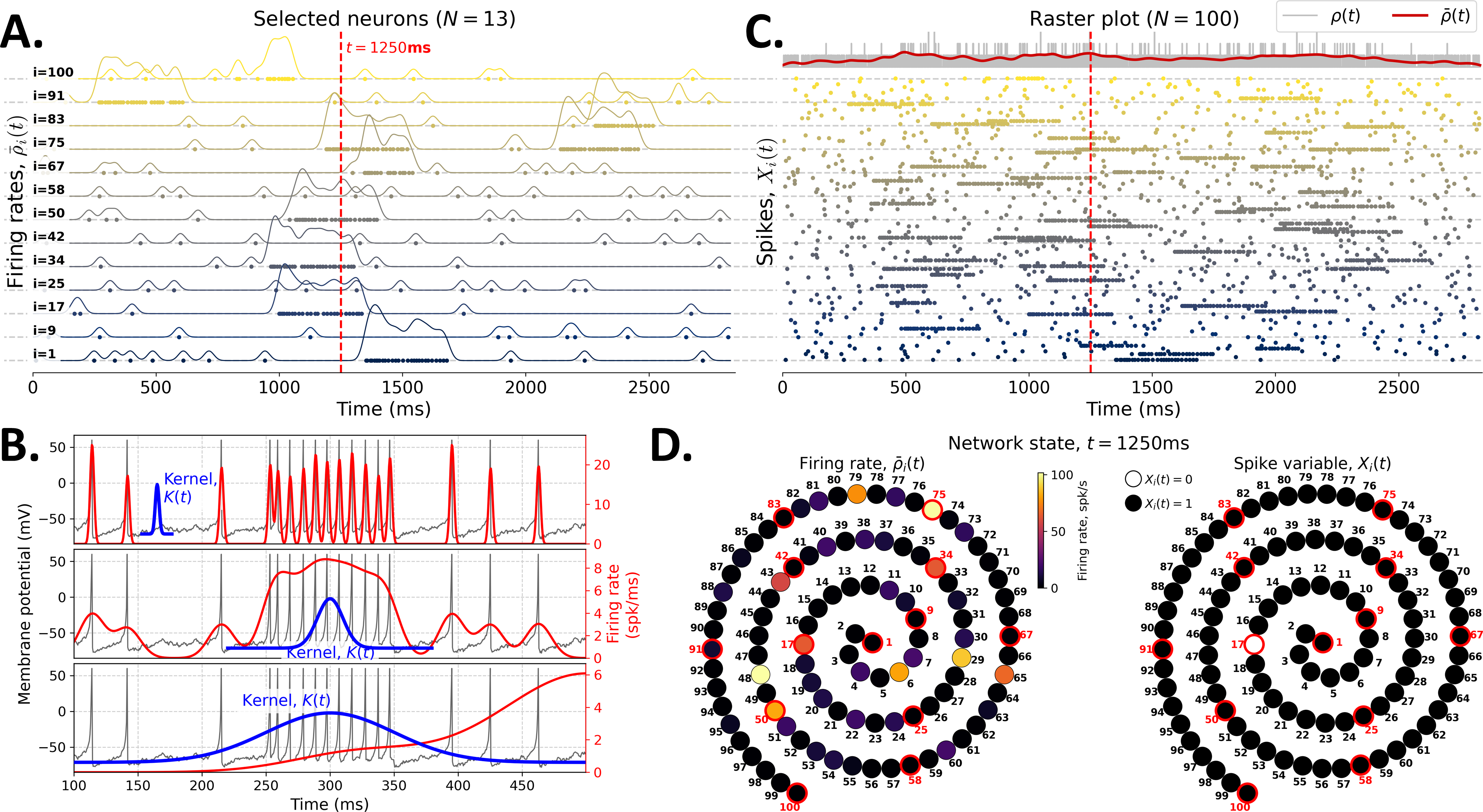}
    \caption{\label{fig:frdef}
\textbf{A.}~Individual firing rates $\bar{\rho}_i(t)$ of the same selected neurons from the simplified simulation shown in Fig.~\ref{fig:dgplacefields}C,D.
\textbf{B.}~Calculation of individual firing rates via convolution with a given Gaussian kernel $K(t)$ (blue curve).
            Top to bottom: increasing the kernel width, there is an optimal value (middle) for which the spiking activity (black curve)
            is more accurately represented by the resulting firing rate (red curve).
\textbf{C.}~Spiking activity (bottom) of all $N=100$ neurons in the simulation, and their population firing rate (top).
\textbf{D.}~Snapshot of the network state at $t=1,250$~ms (vertical dashed lines in panels A,C).
            While raw spiking activity (right) indicates that the network is mostly inactive,
            there is a clear activation pattern present in the firing rates of the population (left).
            For a clear comparison, selected neurons are encircled in red.}
\end{figure*}

To illustrate this idea, consider the simplified simulation shown in Fig.~\ref{fig:dgplacefields}C.
The arena is populated by place fields, each feeding input to an integrate-and-fire (IF)-like neuron.
As the simulated mouse traverses the arena,
neurons whose place fields intersect its trajectory transiently increase their firing activity (Fig.~\ref{fig:dgplacefields}D).
Some neurons exhibit a brief period of elevated firing while the animal crosses their corresponding place field,
Meanwhile, many neurons remain close to their baseline activity because their place fields are never visited.
The increase in firing activity therefore reflects the animal's position (Fig.~\ref{fig:frdef}A).

This behavior naturally motivates a firing-rate description.
Instead of the precise timing of individual spikes,
the variable of interest is the count of spikes within a temporal window.
Neurons whose place fields are currently occupied by the animal exhibit a higher average firing rate.
So, for example, we could imagine that a trajectory elicits a spatiotemporal firing pattern throughout the DG
made of a sequence of neurons temporarillyy increasing their firing rates (see, e.g., Fig.~\ref{fig:frdef}).
This is an example of \textit{rate coding}: information is represented by changes in
the average firing rate of a neuron or neuronal population.
In the next section, we formalize this concept mathematically.

As quick side note, place fields are not fixed: changes in the environment may induce \textit{remapping},
causing their locations to shift~\citep{GoodSmith2017}.
Although the functional role of place fields remains an active topic of research,
they provide a clear example of firing rate activity associated with spatial variables.
We assumed that place fields stand for an actual neural code, but whether it genuinely does coding
or is a position-correlated artifact is an open question.

\subsection{Temporal average}

\subsubsection{Stationary firing rates}

The simplest definition of a neuron's firing rate is obtained by counting
the total number of spikes $n$ recorded over an observation
time $T$ and dividing by the recording duration.
We refer to this quantity as the stationary firing rate, $\rho^*$.
This definition is appropriate for stationary activity,
in which the average firing rate remains approximately constant over time,
\begin{equation}
\label{eq:fr_stationary}
\rho^*=\frac{n}{T}\ .
\end{equation}
In SI units, firing rates are measured in spikes/second, or s$^{-1}$.
In neuroscience, it is common to associate the unit of Hertz, since [Hz]=[s$^{-1}$].
However, this nomenclature is inaccurate, since Hertz is traditionally used
for the frequency of waves --- periodic structures that repeat after a given number of cycles
--- which is physically distinct from counting the number of events (spikes) per unit time.

Although useful in some situations, the definition in Eq.~\eqref{eq:fr_stationary} cannot capture time-dependent signals.
For example, as shown in Fig~\ref{fig:dgplacefields},
when an animal crosses a place field, the synaptic input to a neuron changes over time,
producing transient changes in its firing activity.
Capturing this temporal structure requires a time-dependent definition of the firing rate.

\subsubsection{Capturing temporal fluctuations}

A straightforward extension is to divide the recording into temporal bins
of width $\Delta t$ and count the spikes within each bin. To formalize this procedure,
we first define a binary spike train $X(t)$, which equals one when a spike occurs and zero otherwise,
\begin{equation}
\label{eq:def_spiketrain}
X(t)=\left\{
    \begin{array}{ll}
        1\,,&\textnormal{ if the neuron spiked at time }t\\
        0\,,&\textnormal{ otherwise }
    \end{array}
\right.=\sum\limits_{k=1}^n\delta(t-t_k)\ .
\end{equation}
Here, we use a normalized delta function, $\delta(t-t_k)=1$ if $t=t_k$ (and zero otherwise),
and the sum runs over all observed spikes, indexed by $k$, occurring at times $t_k$.
The binned firing rate $\rho_B$ is then obtained by averaging
the spike train within each temporal window, indexed by $m=1,\ldots,T/\Delta t$,
\begin{equation}
\label{eq:fr_timebinning}
\rho_B(m)=\frac{1}{\Delta t}\sum\limits_{t=(m-1)\Delta t}^{m\Delta t-1}X(t)\ .
\end{equation}
This approach provides a time-dependent estimate of firing activity and forms the basis
of the peristimulus time histogram~\citep{GerstnerBook2014}.
Its main limitation is that the bin width is chosen arbitrarily,
making it primarily a data-analysis tool rather than a model of neural computation.


A smoother estimate is obtained by replacing fixed bins with a sliding temporal window.
Mathematically, this operation corresponds to convolving the spike train with a normalized kernel $K(t)$,
\begin{align}
\label{eq:fr_convolution}
\bar{\rho}(t)=\left(K*X\right)(t)&\equiv\int_{-\infty}^{+\infty}\!\!\!\!\!\!\!\!K(t-s)\,X(s)\,{\rm d}s\\
                                 &=\sum\limits_{k}K(t-t_k)\ ,\nonumber
\end{align}
where we used the spike train from Eq.~\eqref{eq:def_spiketrain} to evaluate the integral.
The overbar indicates that the firing rate $\rho$ is obtained by temporal averaging.
Importantly, this averaging must be applied to the spike train $X(t)$, not to the raw membrane voltage.

The choice of kernel determines the degree of temporal smoothing. Common examples include rectangular, exponential, alpha, and Gaussian kernels,
\begin{align}
\label{eq:conv_rectK}
K_{\rm rect}(t) &= \frac{1}{W}\left[\Theta(t)-\Theta(t-W)\right] && \plotKernelRect \\[1.5ex]
\label{eq:conv_expK}
K_{\rm exp}(t) &=\frac{1}{W}e^{-t/W}\Theta(t) && \plotKernelExp \\[1.5ex]
\label{eq:conv_alphaK}
K_{\alpha}(t) &=\frac{t}{W^2}e^{-t/W}\Theta(t) && \plotKernelAlpha \\[1.5ex]
\label{eq:conv_gaussK}
K_{\rm Gauss}(t) &= \frac{1}{\sqrt{2\pi}W}e^{-t^2/\left(2W^2\right)} && \plotKernelGauss
\end{align}
where $W$ is a width parameter, and $\Theta(t)=1$ for $t\geq0$ (and zero otherwise) is the step function.
The rectangular kernel with $W=\Delta t$ gives a result consistent with Eq.~\eqref{eq:fr_timebinning},
without requiring time to be divided into discrete bins. The exponential kernel captures
the decay of correlations between random events in stochastic processes.
The $K_{\alpha}$ kernel resembles the temporal profile of a postsynaptic potential (PSP)
arriving at a neuron's dendrites. This kernel, also known as the \textit{alpha} function,
is discussed in Box~\ref{box:fr_convolution}~\citep{Roth2010Synaptic}.

Gaussian kernels provide smooth estimates but depend on spikes occurring both before and after the evaluation time $t$.
This kernel was used to calculate the firing rates shown in Fig.~\ref{fig:frdef}.
Figure~\ref{fig:frdef}B illustrates the effect of increasing the Gaussian kernel width $W$.
For narrow kernels, the estimated firing rate closely follows individual spikes.
As the kernel broadens, the signal becomes progressively smoother,
revealing changes in firing activity over time.
However, excessively broad kernels remove temporal details,
illustrating the trade-off between temporal resolution and smoothing.
As the averaging window becomes much longer than the characteristic timescale of fluctuations,
the temporally averaged firing rate should converge to the stationary measure in Eq.~\eqref{eq:fr_stationary}.

From a signal-processing perspective, convolution acts as a low-pass filter,
attenuating rapid fluctuations while preserving slower variations in firing activity.
Calculating a firing rate is therefore generally an irreversible process:
the spike times that generated a given firing rate cannot be recovered precisely.
The appropriate kernel and width $W$ depend on the temporal scale of the phenomenon of interest.
Nevertheless, neuronal populations may implement this type of temporal averaging,
with a suitably tuned $W$ allowing them to transform
temporal fluctuations into a response. See Box~\ref{box:fr_convolution}.

\begin{mauriciofloatingnote}[t!]{Biological interpretation of the temporal average}
    \label{box:fr_convolution}
    \begin{wrapfigure}{l}{0.35\linewidth}
    \vspace{-1.5\baselineskip}
    \centering
    \includegraphics[width=\linewidth]{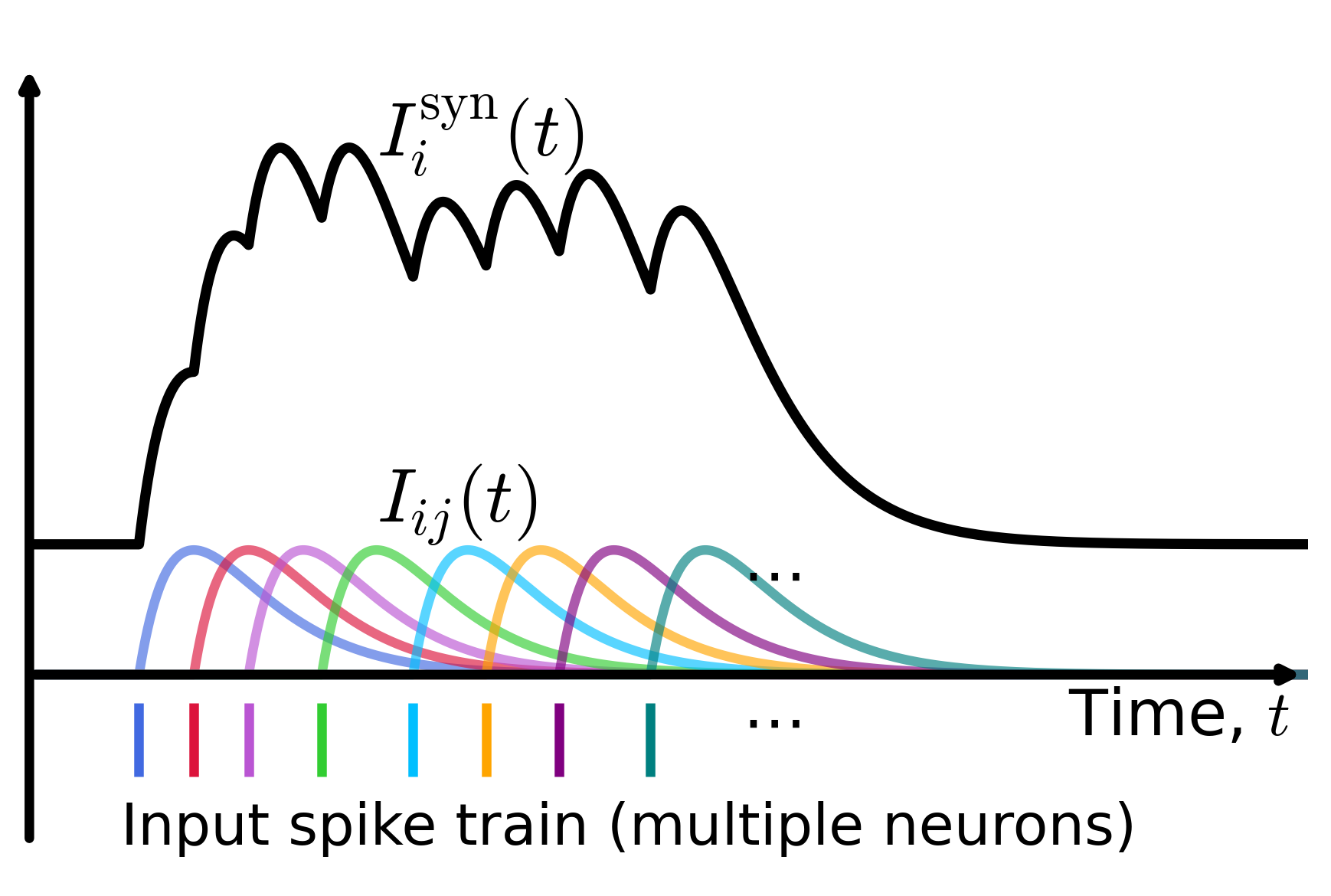}
    \vspace{-\baselineskip}
    \end{wrapfigure}
    Consider a neuron receiving many presynaptic trains of spikes, Eq.~\eqref{eq:def_spiketrain}.
    Each incoming spike index by $k$ generates a PSP approximately in the form
    of an \textit{alpha} function, Eq.~\eqref{eq:conv_alphaK}.
    The sum of the PSPs generated by the spike train $X_j=\sum_k\delta(t-t_{j,k})$
    from presynaptic neuron $j$ is
    \begin{equation}
    \sum_kK_{\alpha}(t-t_{j,k})
            = \int_{-\infty}^{t}\!\!\!\!\! K_{\alpha}(t-s)X_j(s){\rm d}s\equiv \bar{\rho}_j(t)\ ,
    \end{equation}
    from our definition of firing rate, Eq.~\eqref{eq:fr_convolution}. The total synaptic current is
    \begin{equation}
    \label{eq:box_fr_synsum}
    I^{\rm syn}_i(t) = \sum_j G_{ij} \sum_kK_{\alpha}(t-t_{j,k})=\sum_j G_{ij} \bar{\rho}_j(t)\ ,
    \end{equation}
    where $G_{ij}$ are synaptic conductances. Thus, temporal averaging emerges naturally
    from synaptic integration over timescales set by the postsynaptic current dynamics.
    This approximation is relevant when a neuron receives many spikes over several milliseconds.
\end{mauriciofloatingnote}

\subsection{Population firing rates}

The firing rate can also be defined by averaging activity across a population of neurons.
Consider a population of $N$ neurons with spike trains $X_i(t)$ defined in Eq.~\eqref{eq:def_spiketrain}.
At time $t$, the instantaneous population activity is defined as the average of the spike trains,
\begin{equation}
\label{eq:fr_population}
\rho(t)=\avg{X_i(t)}\equiv\frac{1}{N}\sum\limits_{i=1}^{N} X_i(t)=\frac{n(t)}{N}\ ,
\end{equation}
where $n(t)$ is the number of spikes in the population at time $t$.
This quantity represents the average activity across neurons at a given instant,
or equivalently, the average across the columns of a raster plot
(the light gray curve above the raster plot in Fig.~\ref{fig:frdef}C,).

Strictly speaking, however, Eq.~\eqref{eq:fr_population} gives the average
number of spikes per neuron at a given instant, rather than a firing rate per unit time.
To obtain a firing rate, we must also average over time.
Using Eq.~\eqref{eq:fr_convolution} to calculate the temporally
averaged firing rate $\bar{\rho}_i(t)$ of each neuron, we obtain the population firing rate,
\begin{align}
\bar{\rho}(t)=\avg{\bar{\rho}_i(t)}&=\frac{1}{N}\sum_{i=1}^{N}\int_{-\infty}^{+\infty}\!\!\!\!\!\!\!\!K(t-s)\,X_i(s)\,{\rm d}s\nonumber\\
\label{eq:fr_population_time}
                                   &=\int_{-\infty}^{+\infty}\!\!\!\!\!\!\!\!K(t-s)\,\left[\frac{1}{N}\sum_{i=1}^{N}X_i(s)\right]\,{\rm d}s\\
                                   &=\int_{-\infty}^{+\infty}\!\!\!\!\!\!\!\!K(t-s)\,\avg{X_i(s)}\,{\rm d}s
                                    =\int_{-\infty}^{+\infty}\!\!\!\!\!\!\!\!K(t-s)\,\rho(s)\,{\rm d}s\nonumber\ .
\end{align}
Equation~\eqref{eq:fr_population_time} tells us that we can obtain the population firing rate in two equivalent ways.
One is by first calculating the temporally averaged firing rate $\bar{\rho}_i(t)$ of each neuron
and then averaging across the population.
Another is by first averaging the spike trains across neurons to obtain $\rho(t)$
and then averaging over time. In other words, the order of applying Eqs.~\eqref{eq:fr_convolution}
and~\eqref{eq:fr_population} does not matter. Equation~\eqref{eq:fr_population_time} is shown
as a solid red line above the raster plot in Fig.~\ref{fig:frdef}C.
Box~\ref{box:fr_trials} discusses how Eqs.~\eqref{eq:fr_population}
and~\eqref{eq:fr_population_time} can also be interpreted as averages over $N$ experimental trials of the same neuron.

The temporal average is particularly clear when we use the rectangular kernel,
Eq.~\eqref{eq:conv_rectK}, with $W=\Delta t$,
\begin{equation*}
\bar{\rho}(t)=\int\limits_{-\infty}^{+\infty}\frac{1}{\Delta t}\Big[\Theta(t-s)-\Theta(t-s-\Delta t)\Big]\,\rho(s)\,{\rm d}s
             =\frac{1}{\Delta t}\int_{t-\Delta t}^{t}\!\!\!\!\!\!\!\!\rho(s)\,{\rm d}s\ .
\end{equation*}
This follows because $K_{\rm rect}(t-s)\neq0$ only when $0<t-s<\Delta t$,
so $t-\Delta t<s<t$. Thus, $\bar{\rho}(t)\Delta t$ can be interpreted
as the expected number of spikes per neuron during the interval $[t-\Delta t,t]$.
For sufficiently small $\Delta t$, this quantity approximates the probability
of a spike occurring within that interval~\citep{abbottNeuro}.
In other words, the instantaneous rate $\rho(t)$ from Eq.~\eqref{eq:fr_population},
is analogous to a spike probability density for the neuronal population
when $\Delta t$ is of the order of a single spike duration.
These concepts provide the foundation for developing an effective
(mean-field) theory of collective neuronal network behavior.

As the population size $N$ increases, fluctuations in $\rho(t)$ decrease.
For example, if neurons fire independently, the variance of $\rho(t)$
follows the law of large numbers, ${\rm Var}(\rho(t))\sim1/N$,
yielding an increasingly smooth estimate of collective activity.
Correlations between neurons weaken this reduction in fluctuations,
leading to larger variance than expected for independent processes.

Temporal averaging also helps explain why the population-averaged subthreshold membrane
potential can resemble the population firing rate~\citep{Caprioti2026Hopfield}.
Removing action potentials before averaging the membrane voltage traces suppresses these fast events,
revealing slower changes in neuronal activity.
For example, membrane adaptation currents can produce a transient increase
in firing rate after stimulus onset, followed by a gradual relaxation toward a lower steady-state level,
similarly to what is observed in membrane potentials.

\begin{mauriciofloatingnote}[t!]{Trial averaging}
    \label{box:fr_trials}
    \begin{wrapfigure}{l}{0.5\linewidth}
    \vspace{-1\baselineskip}
    \centering
    \includegraphics[width=\linewidth]{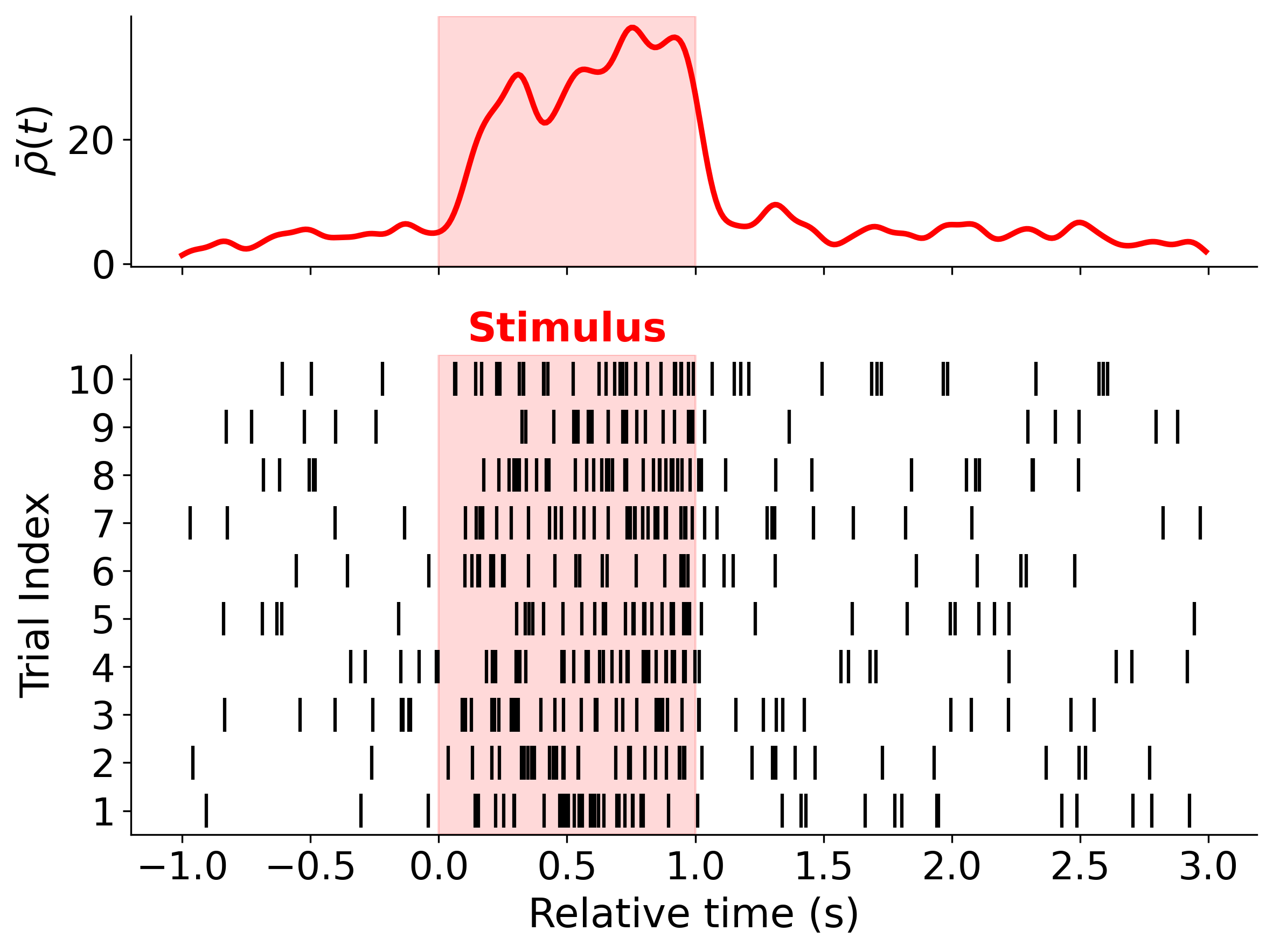}
    \vspace{-\baselineskip}
    \end{wrapfigure}
A final definition of firing rate is commonly used in experimental neuroscience.
Instead of averaging over time or across a population of neurons,
activity is averaged across repeated presentations of the same experimental protocol (trials).
The definition is identical to that of the population firing rate,
Eqs.~\eqref{eq:fr_population} and~\eqref{eq:fr_population_time}, except that $i$ denotes the trial index.
Here, we used a Gaussian kernel with $W=50$~ms.

This approach assumes that the same neuron is recorded across repeated
presentations of an identical stimulus or behavioral condition.
Averaging across trials suppresses trial-to-trial variability
and reveals the reproducible component of the neuronal response.
Unlike population averaging, however, trial averaging does not necessarily
represent a computation performed by the nervous system.
Instead, it is an experimental tool for characterizing
neuronal responses and comparing them across conditions.

These definitions of firing rate are closely related and provide complementary
descriptions of neuronal activity. Depending on the application,
one may be more convenient than another for theoretical analysis
or the interpretation of experimental data.
\end{mauriciofloatingnote}

\section{Coding with firing rates}

Neurons continuously transform incoming signals into outgoing activity.
More generally, any process that maps an input signal to an output signal
can be regarded as a coding scheme~\citep{OlshausenSparseCode2004}.
The nervous system uses several coding strategies, including temporal,
population, and phase coding~\citep{abbottNeuro,GerstnerBook2014}.
More recently, variance has also been proposed as a means of encoding inputs~\citep{Girardi2025RateVar}.
Here, motivated by the problem of place fields, we focus on \textit{rate coding},
in which information is represented by the firing rate of individual neurons or neuronal populations.

\begin{figure*}[t!]
    \includegraphics[width=1\linewidth]{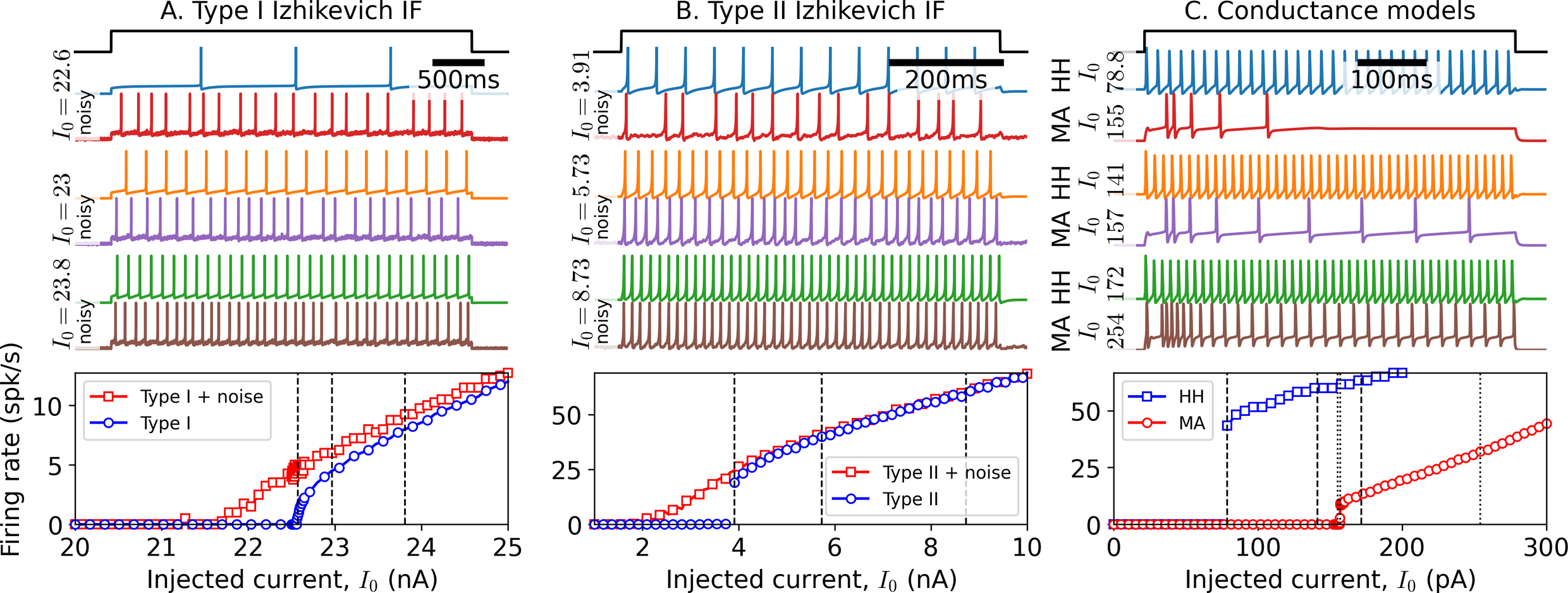}
    \caption{\label{fig:ficurves}$F$-$I$ curves (bottom) and selected voltage traces (top)
            generated from DC current injections of selected intensities (vertical lines below).
            Neurons are only activated when input exceeds a given threshold $I_c$~\citep{izhikevichDyn}.
\textbf{A.}~A type I membrane transitions into spiking with arbitrarily small firing rate.
\textbf{B.}~A type II membrane abruptly goes from no spikes to a well-defined firing rate.
\textbf{A and B.} Weak noise smoothens the transition, generating spikes for subthreshold currents
                  and hiding the type II all-or-none characteristic.
\textbf{C.}~Conductance-based models exhibit the same behavior. The classical HH neuron presents a
            type II transition~\citep{Ermentrout1996TypeI}; Adding adaptive mechanisms, such as an M-type K\textsuperscript{+} current,
            can also smoothen the transition and generate spike frequency adaptation~\citep{Gutkin2014SpikeFreqA}.}
\end{figure*}

\subsection{Single neuron: the F-I curve}
\label{sec:f-Icurve}

For an isolated neuron, the current flowing through the soma membrane can be described by~\citep{izhikevichDyn},
\begin{equation}
\label{eq:membranevolt}
C \frac{{\rm d}V}{{\rm d}t}=f(V)+I(t),
\end{equation}
where $C$ is the membrane capacitance, $V$ is the membrane potential,
$f(V)$ describes the intrinsic processes governing membrane
dynamics, and $I(t)$ is the net current arising from synaptic inputs or external sources.
The dynamics described by Eq.~\eqref{eq:membranevolt}
also requires a reset rule to generate repetitive spikes,
since it only describes the dynamics up to spike initiation.
This class of models is called \textit{integrate-and-fire} (IF)~\citep{izhikevichDyn}.
In many neurons, the nonlinear dynamics described by $f(V)$ give rise to excitability:
spikes are generated when a constant injected current exceeds an activation
threshold $I_c$, i.e., $I(t)>I_c$.

\cite{hodgkinClasse} classified neuronal excitability based on the relationship
between injected current and the resulting firing response.
The firing rate versus input current ($F$-$I$) curve is obtained by applying
long-lasting constant (DC) current steps of different amplitudes.
For each current value ($I(t)\sim I_0$), the neuron is allowed to reach a steady firing regime.
The initial transient is then discarded, and the firing rate is calculated from
the remaining spikes using Eq.~\eqref{eq:fr_stationary} over a sufficiently long interval $T$.
Repeating this procedure for different current amplitudes yields the $F$-$I$ curve.

This protocol distinguishes three types of neuronal excitability.
Type I neurons can fire at arbitrarily low rates as $I_0\to I_c$ from above,
producing a continuous onset of firing in the $F$-$I$ curve (Fig.~\ref{fig:ficurves}A).
Type II neurons begin firing at a finite, nonzero frequency when $I_0=I_c$,
and their firing rates typically vary over a relatively narrow range
as the injected current changes (Fig.~\ref{fig:ficurves}B).
Type III neurons produce only a few transient spikes (not shown),
even in response to moderately strong stimuli,
making them unsuitable for sustained rate coding.

\begin{mauriciofloatingnote}[t!]{$F$-$I$ curves and bifurcations}
    \label{box:bifurcation}
Although the shape of the $F$-$I$ curve provides insight into
the transition from rest to repetitive spiking,
it is a phenomenological description rather
than a unique signature of the underlying membrane dynamics or the bifurcation at spike onset.
Different dynamical mechanisms can produce similar $F$-$I$ curves,
while spike shape, adaptation currents acting on different timescales,
and membrane noise can also influence the measured firing rate (Fig.~\ref{fig:ficurves}).
Here, however, we focus on the $F$-$I$ curve as an experimentally measurable
relationship between input intensity and firing rate. Some neurons, known as resonators,
respond preferentially to input frequency rather than intensity,
whereas others produce only a few transient spikes regardless of stimulus duration.
We therefore do not consider the specific membrane dynamics encoded in the function
$f(V)$ in Eq.~\eqref{eq:membranevolt}. A detailed account
of bifurcations underlying spike onset is given by \cite{izhikevichDyn}.
\end{mauriciofloatingnote}

Many neurons exhibit spike-frequency adaptation: their firing rate gradually
decreases during sustained current injection, even though the input remains constant (Fig.~\ref{fig:ficurves}C).
This behavior can arise from slowly activating ionic currents
that reduce neuronal excitability, such as the M-type potassium current~\citep{Gutkin2014SpikeFreqA}.
In contrast, faster adaptation mechanisms, such as A-type currents,
tend to delay spike initiation~\citep{Geiger2000Acurr}. More generally,
the intrinsic membrane dynamics represented by $f(V)$ in Eq.~\eqref{eq:membranevolt}
also depend on gating variables that describe ion-channel kinetics
and other slow membrane processes, which were omitted here for simplicity.
A summary of the main ionic currents is given by \cite{izhikevichDyn}.

Figure~\ref{fig:ficurves}C illustrates the combined effects of A- and M-type currents:
the first spike is delayed after stimulus onset, and the firing rate gradually
decreases during sustained stimulation. Some of these currents effectively
raise the spike threshold over time~\citep{Benda2021Adapt},
slowing the firing rate while reducing noise and improving information transfer~\citep{Chacron2007}.
More broadly, adaptation can support gain control, temporal filtering,
efficient coding~\citep{Benda2021Adapt},
and coding through firing-rate variance~\citep{Girardi2025RateVar}.

The type of neuronal excitability also has important computational implications.
Type I neurons, exemplified by some cortical excitatory pyramidal neurons,
can map input strength continuously to firing rate, making them well suited
for analog rate coding. Type II neurons, such as fast-spiking cortical
inhibitory interneurons, typically begin firing at a finite frequency
and may operate over a more restricted frequency range.
They can contribute to precise temporal coordination
and network synchronization~\citep{Ermentrout1996TypeI},
as well as selectively filter inputs according to their properties.
Type III neurons do not sustain repetitive firing under
typical conditions and are instead associated with transient responses or coincidence detection.

Membrane noise can arise from randomly arriving subthreshold synaptic
currents~\citep{Trinh2023hMC} and can facilitate information transfer in excitable systems.
The accumulation of these weak voltage increments can trigger spikes even when
the coding input remains below $I_c$, enabling neurons to detect subthreshold stimuli.
This effect can smooth the transition between inactive and active regimes
in both Type I and Type II neurons (Fig.~\ref{fig:ficurves}A and~B).

Despite the diversity of membrane dynamics, tonically spiking neurons
can often be approximated by a transfer function (the $F$-$I$ curve)
that relates input strength to output firing rate~\citep{Benda2010Thresh}.
This curve therefore provides another building block for firing-rate models.
However, this description has two important limitations:
\begin{itemize}
\item It assumes that the relevant information is encoded in input strength,
      accumulated over time, rather than in the precise timing of individual inputs.
\item It assumes that both the neuron and the downstream readout
      system integrate activity over hundreds of milliseconds.
\end{itemize}
Consequently, rate coding is appropriate when the relevant
timescales are much longer than those of the underlying membrane dynamics.

Under these conditions, the $F$-$I$ curve can simplify the description of
a tonically spiking neuron~\citep{Roxin2011DistFR}.
As explained in Box~\ref{box:fr_trials}, Eq.~\eqref{eq:fr_population}
can be interpreted as the average spike train over $N$ trials
in which the same neuron receives a stationary stimulus $I(t)$. Thus,
\begin{equation}
\label{eq:XavgF}
    \avg{X_i(t)}\equiv F(I(t))\ .
\end{equation}
This equality implicitly assumes ergodicity:
the average spike train across repeated trials is equivalent
to the time average used to construct the $F$-$I$ curve.
This follows when the $F$-$I$ curve is enough to
determine the input-output relation of the membrane,
as illustrated in Fig.~\ref{fig:ficurves}.

When neuronal activity is temporally averaged using an exponential kernel,
Eq.~\eqref{eq:conv_expK}, with $W=\tau$, we can derive the dynamics of the firing rate.
Here, $W$ sets the timescale of temporal averaging and is assumed
to be comparable to the combined duration $\tau$ of the spike and its refractory period.
Substituting the exponential kernel into Eq.~\eqref{eq:fr_population_time},
together with Eq.~\eqref{eq:XavgF}, and taking the time derivative gives
\begin{align*}
    \frac{{\rm d}\bar{\rho}}{{\rm d}t}
    &=\int\limits_{-\infty}^{+\infty}\frac{{\rm d}}{{\rm d}t}K(t-s)\,F(I(s))\,{\rm d}s\\
    &=\frac{1}{\tau}\int\limits_{-\infty}^{+\infty}
    \left[-K(t-s)+e^{-(t-s)/\tau}\delta(t-s)\right]F(I(s))\,{\rm d}s,
\end{align*}
where we used ${\rm d}K/{\rm d}t=\left[-K(t)+e^{-t/\tau}\delta(t)\right]/\tau$.
Simplifying, we obtain
\begin{equation}
\label{eq:fr_dynamics_neuron}
\tau\frac{{\rm d}\bar{\rho}}{{\rm d}t}=-\bar{\rho}+F(I(t))
\end{equation}
which defines a rate-based model of a neuron.
The parameter $\tau$ sets the timescale over which the
firing rate responds to changes in input. This approximation
is most appropriate when $\tau$ is much shorter than the characteristic
timescales of the phenomena being modeled. For a derivation
from a statistical-physics perspective, see \cite{Ermentrout2010Book,peretto}.

The derivation does not depend on the source of the input,
only on its time scale, so $I(t)$ can represent
a combination of slowly varying synaptic inputs and external stimuli.
In a network of $N$ neurons,
each neuron $i$ has a firing rate $\bar{\rho}_i$ governed by Eq.~\eqref{eq:fr_dynamics_neuron}.
As discussed in Box~\ref{box:fr_convolution}, the total input to neuron $i$
can be decomposed into external and synaptic contributions:
\begin{equation}
\label{eq:inputrate}
I_i(t)=I^{\rm ext}_i(t) + I^{\rm syn}_i(t)=I^{\rm ext}_i(t) + \sum_j G_{ij}\bar{\rho}_j(t)\ ,
\end{equation}
where the sum runs over all presynaptic neurons connected to neuron $i$.

For example, \cite{Hopfield1982,Hopfield1984} studied pattern matching and
content-addressable memory using rate-based neurons governed by equations
of the form of Eqs.~\eqref{eq:fr_dynamics_neuron} and~\eqref{eq:inputrate},
in both continuous and discrete time. Hopfield showed that interactions
among rate-based units can store memories in the network connectivity, $G_{ij}$,
and retrieve them from incomplete patterns. For this work,
he shared the 2024 Nobel Prize in Physics~\citep{Nobel2024}.
More broadly, rate-based units of this kind have played a
central role in neural network theory, including in models
that underpin modern artificial intelligence~\citep{hertzRedesNeurais,Rojas1996,peretto,Girardi2013map,Caprioti2026Hopfield}.

\subsection{Populations of neurons}

Population rate coding can operate faster than single-neuron rate
coding because temporal averaging is not required to define
a population activity measure, as discussed in Eq.~\eqref{eq:fr_population}.
The collective activity of many neurons defines an instantaneous population firing rate,
allowing population interactions to transmit rate-coded signals over shorter timescales
than those typically needed to estimate the firing rate of an individual neuron.

At the population level, the input is the average firing rate received by the population,
and the output is the average firing rate it generates.
The resulting population transfer function depends not only on the intrinsic dynamics
of individual neurons but also on collective network properties,
including the synaptic coupling strengths $G_{ij}$ and synaptic dynamics.

Consequently, a network's input-output relationship is not simply
the sum of the transfer functions of its constituent neurons.
Recurrent interactions reshape the population transfer function,
modifying its gain, dynamic range, and sensitivity to external stimuli.
In particular, networks operating near critical regimes have been shown
to exhibit enhanced sensitivity to weak inputs~\citep{kinouchiCopelli,Galera2020,Girardi2025RateVar}.
In such networks, weak inputs can produce distinguishable stronger outputs than
their noncritical counterparts. This enhancement of coding emerges
from collective network dynamics rather
than from the properties of individual neurons alone.
We will show this effect in the following subsections.

\subsubsection{From spikes to rates: mean-field approximation}
\label{sec:meanfield}

We now derive a firing-rate equation for a population of $N$ neurons and then discuss its rate-coding capacity.
Since we are interested in spike generation, we begin with the membrane dynamics in Eq.~\eqref{eq:membranevolt},
expressed on the timescale of neuronal spiking. Multiplying the membrane equation by the membrane
resistance $R$ introduces the membrane time constant $\tau=RC$. We then discretize
the equation using the Euler method with time step $\Delta t$, measuring time in units of $\Delta t$. This gives
\begin{equation*}
\tau_i \frac{{\rm d}V_i}{{\rm d}t}
    \approx \tau_i \frac{V_i((t+1)\Delta t)-V_i(t\,\Delta t)}{\Delta t}
    =R_if(V_i(t\,\Delta t))+R_i I_i(t\,\Delta t)\ ,
\end{equation*}
where the index $i$ identifies the variables and parameters of neuron $i$ in the network.

We can make the leak current explicit by writing
$R_if(V_i)=-V_i+V^b_i+f_0(V_i)$, where $V^b_i$ controls the resting potential of neuron $i$,
and $f_0(V_i)$ captures nonlinear contributions. We then omit $\Delta t$
from the arguments of $V_i$ and $I_i$, using it as the unit of time.
Finally, we rescale the input current by the membrane resistance,
absorbing the factor $R_i$ into its definition, so that $R_iI_i\to I_i$.
The resulting equation is
\begin{equation}
\label{eq:Veuler}
V_i(t+1)=\left(1-\frac{\Delta t}{\tau_i}\right)V_i(t)+\frac{\Delta t}{\tau_i}\left[V^b_i+f_0(V_i(t))+I_i(t)\right]\ .
\end{equation}

The discrete-time description assumes that the network evolves on a
timescale much slower than that of individual neurons. To formalize
this separation of timescales, we assume that neurons fire stochastically.
If $\avg{\tau_i}=\tau$ is the average interval between firing events,
then $N\tau$ is the characteristic time associated with $N$ firing events
in a network of $N\gg1$ neurons. We use this separation of timescales
to choose an integration step $\Delta t\lesssim\tau$, such that
$\Delta t/\tau_i\approx1$ for the neurons of interest. This approximation
focuses on the long-time network dynamics rather than the detailed
evolution of individual membrane potentials. Together with a
firing-and-reset rule, the neuronal dynamics can then be approximated by
\begin{equation}
\label{eq:GLmodel}
V_i(t+1)=\left\{
\begin{array}{ll}
    0                        &\textnormal{ if } X_i(t)=1\ ,\\
    V^b_i+I_i(t) &\textnormal{ otherwise, }
\end{array}
\right.
\end{equation}
where $X_i(t)=1$ if neuron $i$ fires at time $t$, and $X_i(t)=0$
otherwise. Here, we have set $f_0(V_i)=0$. 
The input current includes both external and synaptic contributions,
$I_i(t)=I^{\rm ext}_i(t) + I^{\rm syn}_i(t)$. A typical choice for the synaptic input is
\[
I^{\rm syn}_i(t)=\frac{1}{N}\sum_j G_{ij} X_j(t)\ ,
\]
where the sum runs over the presynaptic neurons connected to neuron $i$.

The stochastic assumption means that the precise timing of individual spikes is not important,
so firing can be described probabilistically.
We define the probability that neuron $i$ fires at time $t$,
given its membrane potential $V_i\equiv V_i(t)$ from Eq.~\eqref{eq:GLmodel},
as $\mathcal{P}\left(X_i(t)=1|V_i\right)\equiv\Phi(V_i)$.
The function $\Phi(V_i)$, called the firing function,
can take various forms, including sigmoidal functions
such as the hyperbolic tangent or logistic function.
A simple example is
\begin{equation}
\label{eq:phifuncGL}
\Phi(V_i)=\left\{
\begin{array}{ll}
0                        &\textnormal{ if } V_i<\theta_i\ ,\\
\Gamma_i\,(V_i-\theta_i) &\textnormal{ if } \theta_i\leq V_i < \theta_i+1/\Gamma_i\ ,\\
1                        &\textnormal{ otherwise. }
\end{array}
\right.
\plotGLPhiFunc
\end{equation}
Here, $\theta_i$ is an absolute firing threshold:
the neuron cannot fire when $V_i<\theta_i$.
Above this threshold, the firing probability increases linearly with $V_i$
until it reaches one at $V_i=\theta_i+1/\Gamma_i$. The parameter $\Gamma_i$
controls neuronal excitability: larger values produce a higher firing
probability at a given potential above threshold. Because $V_i$ evolves
according to Eq.~\eqref{eq:GLmodel}, the firing probability changes over time.
Thus, although $\theta_i$ is fixed, the actual spiking threshold fluctuates,
since a spike can happen with any potential between $\theta_i$ and $\theta_i+1/\Gamma_i$.
A version of this model was first introduced by \cite{Gerstner1992AssocMem} and later
formalized as a generic stochastic IF model by \cite{galvesLocherbach2013}.

The mean firing rate of the network is defined as
$\rho(t)=\avg{X_i(t)}$. However, since the firing variable $X_i(t)$
depends on the membrane potential $V_i(t)$, the average firing rate
must account for the distribution $\mathcal{P}_t(V)$ of membrane
potentials across neurons at time $t$. Thus, taking the average
over the potentials at time $t$ from Eq.~\eqref{eq:GLmodel},
we get the firing rate at step $t+1$,
\begin{align}
\rho(t+1) &= \int_0^\infty\!\!\!\!\! \mathcal{P}(X=1|V) \:\mathcal{P}_t(V) \: dV\nonumber\\
          &= \int_0^\infty\!\!\!\!\! \Phi(V) \:\mathcal{P}_t(V) \: dV\ .
    \label{eq:fr_GLmodel}
\end{align}

The mean-field approximation assumes that the network consists of two
groups of neurons: an active fraction $\rho(t)$, whose membrane
potentials are reset to $V=0$ according to Eq.~\eqref{eq:GLmodel},
and an inactive fraction $1-\rho(t)$, all assigned the same membrane
potential $\avg{V_i(t)}$. Here, $\avg{V_i(t)}$ denotes the average
membrane potential across the neurons that have $X_i(t)=0$ in Eq.~\eqref{eq:GLmodel}.
This approximation replaces
the full distribution of membrane potentials with two delta peaks:
\begin{equation}
\label{eq:GLmodel_PofV}
\mathcal{P}_t(V) = \rho(t)\,\delta(V)+\left[1-\rho(t)\right]\,\delta(V - \avg{V_i(t)})\ .
\end{equation}
Substituting Eqs.~\eqref{eq:GLmodel}, \eqref{eq:phifuncGL}, and
\eqref{eq:GLmodel_PofV} into Eq.~\eqref{eq:fr_GLmodel} yields the
mean-field equation for the firing rate,
\begin{align}
\rho(t+1)&=\left[1-\rho(t)\right]\Phi(\avg{V_i(t)})\nonumber\\
         &=\left[1-\rho(t)\right]\Phi(V^b+I(t))\ ,
\label{eq:GLmodel_MF}
\end{align}
where $V^b\equiv\avg{V^b_i}$ and $I(t)=\avg{I_i(t)}$.
This result assumes that $\Phi(0)=0$, so a neuron cannot fire
immediately after its membrane potential has been reset. The model
therefore has an absolute refractory period of one time step. Although
this condition is not required in general, allowing $\Phi(0)\neq0$
would introduce an additional term in Eq.~\eqref{eq:GLmodel_MF}.

The derivation can be extended to include membrane recovery at a rate
$\mu_i$, by introducing a term $-\mu_iV_i(t)$ in Eq.~\eqref{eq:GLmodel}.
The distribution in Eq.~\eqref{eq:GLmodel_PofV} must then be modified
to account for the resulting spread in membrane potentials.
See \cite{brochiniPhasTrans2016,Kinouchi2019,Girardi2020bal,Girardi2021theory}
for further details and extensions.

We can rewrite Eq.~\eqref{eq:GLmodel_MF} in a more familiar form by
returning to continuous time. Recall that the discrete time index
represents intervals of duration $\Delta t$, with $\Delta t\approx\tau$.
Subtracting $\rho(t)$ from both sides of Eq.~\eqref{eq:GLmodel_MF}
and approximating the finite difference by a time derivative gives
\begin{equation}
\label{eq:GLmodel_conttime}
\tau\frac{\rho((t+1)\Delta t)-\rho(t\Delta t)}{\Delta t}
\approx \tau \frac{{\rm d}\rho}{{\rm d}t}
=-\rho+\left[1-\rho\right]\Phi(V^b+I)\ ,
\end{equation}
where $I$ combines the mean external and synaptic inputs. This equation
resembles Eq.~\eqref{eq:fr_dynamics_neuron}, which describes the firing
rate of an isolated neuron averaged over time and trials. In both cases, the
functions $\Phi$ and $F$ act as nonlinear filters of the net input
received by the neurons. The prefactor $1-\rho$ captures the refractory
period of 1 time step that is implied in the choice of $\Phi(0)=0$,
so that neurons cannot fire immediately after the reset to $V=0$.
Thus, equation~\eqref{eq:fr_dynamics_neuron} can be seen as modeling
the rate of a population of neurons without refractory period.

As an application of this formalism, consider a recurrent network
composed of $\NE$ excitatory neurons (superscript E) and $\NI$
inhibitory neurons (superscript I), as shown in Fig.~\ref{fig:ratecoding}A.
Each neuron follows Eq.~\eqref{eq:GLmodel}, with membrane potentials
evolving according to
\begin{equation}
\label{eq:GLmodelEI}
\begin{array}{rl}
\VE_i(t+1) &=\displaystyle \VbE_i + \IextE[i](t)
+ \dfrac{1}{N}\sum_{j\neq i\in {\rm E}}\WEE_{ij}\XE_j(t)
- \dfrac{1}{N}\sum_{j\in {\rm I}}\WEI_{ij}\XI_j(t)\ ,\\[1em]
\VI_i(t+1) &=\displaystyle \VbI_i + \IextI[i](t)
+ \dfrac{1}{N}\sum_{j\in {\rm E}}\WIE_{ij}\XE_j(t)
- \dfrac{1}{N}\sum_{j\neq i\in {\rm I}}\WII_{ij}\XI_j(t)\ ,
\end{array}
\end{equation}
where excitatory and inhibitory spikes, $\XE_i(t)=1$ and $\XI_i(t)=1$,
occur with probabilities $\PhiE(V)$ and $\PhiI(V)$, respectively.
Each spike resets the corresponding membrane potential to zero at the
next time step.

The population-averaged membrane potentials are
\[
\avg{\VE_i}=\frac{1}{\NE}\sum_{i\in {\rm E}} \VE_i
\qquad\textnormal{and}\qquad
\avg{\VI_i}=\frac{1}{\NI}\sum_{i\in {\rm I}} \VI_i\ .
\]
Applying the mean-field approximation to each population gives
\begin{equation}
\label{eq:fr_GLmodelEI}
\begin{array}{rl}
\rhoE(t+1) &=\displaystyle [1-\rhoE(t)]\,
\PhiE\!\left(\VbE+\IextE(t)+p\,\WEE\,\rhoE(t)
-q\,\WEI\,\rhoI(t)\right)\ ,\\[1em]
\rhoI(t+1) &=\displaystyle [1-\rhoI(t)]\,
\PhiI\!\left(\VbI+\IextI(t)+p\,\WIE\,\rhoE(t)
-q\,\WII\,\rhoI(t)\right)\ ,
\end{array}
\end{equation}
where $p=\NE/N$ and $q=1-p=\NI/N$ are the fractions of excitatory
and inhibitory neurons, respectively. The remaining parameters are
population averages:
\[
\VbE=\frac{1}{\NE}\sum_{i\in {\rm E}} \VbE_i
\qquad\textnormal{and}\qquad
\VbI=\frac{1}{\NI}\sum_{i\in {\rm I}} \VbI_i\ ,
\]
and
\[
\IextE(t)=\frac{1}{\NE}\sum_{i\in {\rm E}} \IextE[i](t)
\qquad\textnormal{and}\qquad
\IextI(t)=\frac{1}{\NI}\sum_{i\in {\rm I}} \IextI[i](t)\ .
\]
Deriving these coupled population equations also requires a homogeneity
assumption on the synaptic couplings. Specifically, for each pair of
populations A and B, the mean coupling from a presynaptic neuron
$j\in{\rm B}$ to the postsynaptic population A is assumed to be
independent of the identity of neuron $j$. We define this average as
\[
G^{\rm AB}=\frac{1}{N^{\rm A}}\sum_{i\in {\rm A}}G^{\rm AB}_{ij}\ ,
\]
and assume that it takes the same value for every presynaptic neuron
$j\in{\rm B}$.

When tuned near the critical point, the E-I model described above
was shown to reproduce fluctuations in cortical activity recorded
from urethane-anesthetized rats \citep{Carvalho2021}.
It also exhibits synaptic balance \citep{Girardi2020bal}.
Using Eq.~\eqref{eq:GLmodel_conttime} to reverse Eq.~\eqref{eq:fr_GLmodelEI} to continuous time
yields the Wilson--Cowan model for E-I networks with non-zero refractory period \citep{WilsonCowan1972}.
Wilson--Cowan models have been used to describe a wide range
of phenomena, from binocular rivalry and decision making to epilepsy
\citep{Cowan2016WC,WilsonCowan2021}.
As discussed above, rate-based equations are well suited to modeling
phenomena that unfold over timescales of hundreds of milliseconds.
Other mean-field approaches can also yield these equations without
relying on time discretization; see, for example,
\cite{Sejnowski1976Stochastic,Ermentrout2010Book,Montbrio2015MFQIF}.

\subsubsection{Rate coding and criticality}

\begin{figure*}[t!]
    \includegraphics[width=1\linewidth]{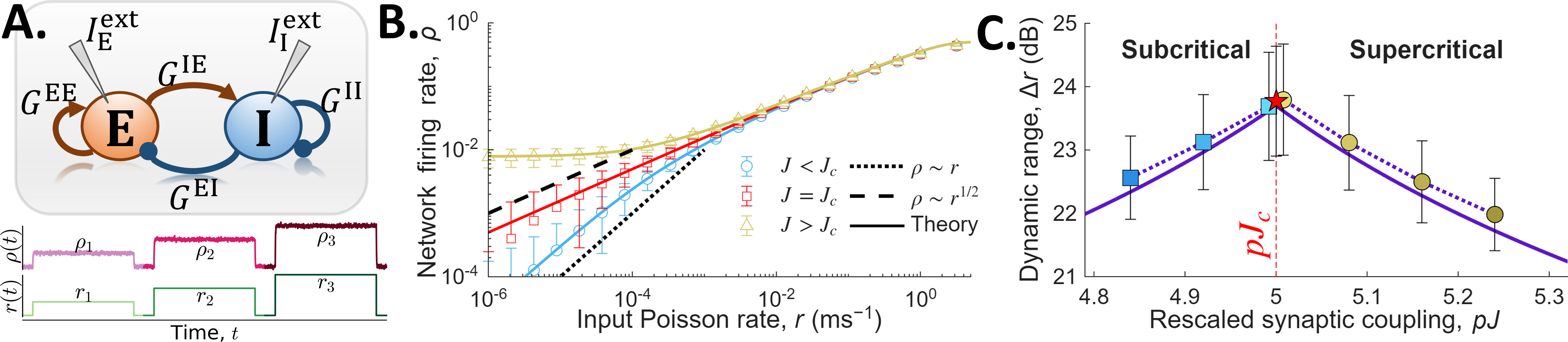}
    \caption{\label{fig:ratecoding}
\textbf{A, top.}~Mean-field E-I network from Eq.~\eqref{eq:fr_GLmodelEI}.
\textbf{A, bottom.}~Simulation scheme: we sequentially inject inputs of rate $r$, and measure
        the corresponding network firing rate $\rho$ at the stationary state.
\textbf{B.}~Network output firing rate as a function of the input rate.
        The shape of the curve depends on network parameters such as the coupling strength, $J$.
        Only at $J=J_c=5/p$ is the network capable of capturing weak stimuli
        (e.g, below $r=10^{-4}$ spikes/ms for each excitatory neuron).
\textbf{C.}~Dynamic range of the rate curves in panel B.
        The maximum dynamic range (rate-coding capacity) happens at the critical point, $J_c=5/p$.
        Solid curves: mean-field theory, Eq.~\eqref{eq:fr_GLmodelEI};
        symbols: simulation of $N=10^5$ neurons following Eqs.~\eqref{eq:GLmodelEI}, \eqref{eq:phifuncTotalGL} and~\eqref{eq:phifuncGL}.}
\end{figure*}

To understand how neuronal populations can encode inputs differently
from isolated neurons, let us introduce an external stimulus to the
network described by Eq.~\eqref{eq:GLmodelEI}.

Consider a Poisson process with rate $r$ spikes/ms/neuron. The
probability that a neuron receives at least one external spike during
a time step of $1$ ms is $P(r)=1-e^{-r}$. We can incorporate this
stochastic input directly into the firing function of the excitatory
population. At each time step, an excitatory neuron can fire either
in response to its synaptic inputs, with probability $\Phi(V)$, or
in response to the external process, with probability $P(r)$.
Combining these two independent activation mechanisms gives
\begin{equation}
\label{eq:phifuncTotalGL}
\PhiE(V)=\Phi(V)+P(r)\delta(V)-\Phi(V)P(r)\ ,
\end{equation}
where $\Phi$ is the piecewise-linear firing function defined in
Eq.~\eqref{eq:phifuncGL}, with firing gain $\Gamma=1/5$ and threshold
$\theta=1$. For simplicity, we set $\PhiI=\Phi$. The factor
$\delta(V)$ ensures that external input cannot activate a neuron
immediately after its membrane potential has been reset, thereby
preserving the one-step refractory period. We then set
$\WEE=\WIE=J$, $\WEI=gJ$, and $\WII=G$ to control the synaptic
couplings with a smaller set of parameters (Fig.~\ref{fig:ratecoding}A, top).

To characterize the input-output relationship, we apply an input
with a fixed rate $r$ over a long interval and measure the stationary
output rate $\rho$ (Fig.~\ref{fig:ratecoding}A, bottom). Repeating
this procedure for different input rates yields the response curve
$\rho(r)$ shown in Fig.~\ref{fig:ratecoding}B.

Consider first weak inputs, $r\ll1$ spike/ms/neuron. As the coupling
strength $J$ crosses the critical value
$J_c=1/(\Gamma p)=5/p$, the input-output relation changes qualitatively.
For $J<J_c$, the response is linear, $\rho\sim r$; at $J=J_c$, it
follows a square-root law, $\rho\sim r^{1/2}$; and for $J>J_c$, the
response curve flattens as the input becomes weak. These response
laws follow from the stationary solutions of
Eq.~\eqref{eq:fr_GLmodelEI}, obtained by setting
$\rhoE(t+1)=\rhoE(t)\equiv\rhoE^*$ and
$\rhoI(t+1)=\rhoI(t)\equiv\rhoI^*$. The corresponding average network
rate is $\rho=p\rhoE^*+q\rhoI^*$. Thus, the network's input-output
relation can change dramatically through collective neuronal dynamics,
controlled here by the synaptic strength $J$. This mechanism has been
proposed as a basis for Stevens' law in psychophysics
\citep{copelliRoqueExcMedia,kinouchiCopelli}.

\subsection{Tuning curves}
\label{sec:tuningcurve}

\begin{figure*}[t!]
    \includegraphics[width=1\linewidth]{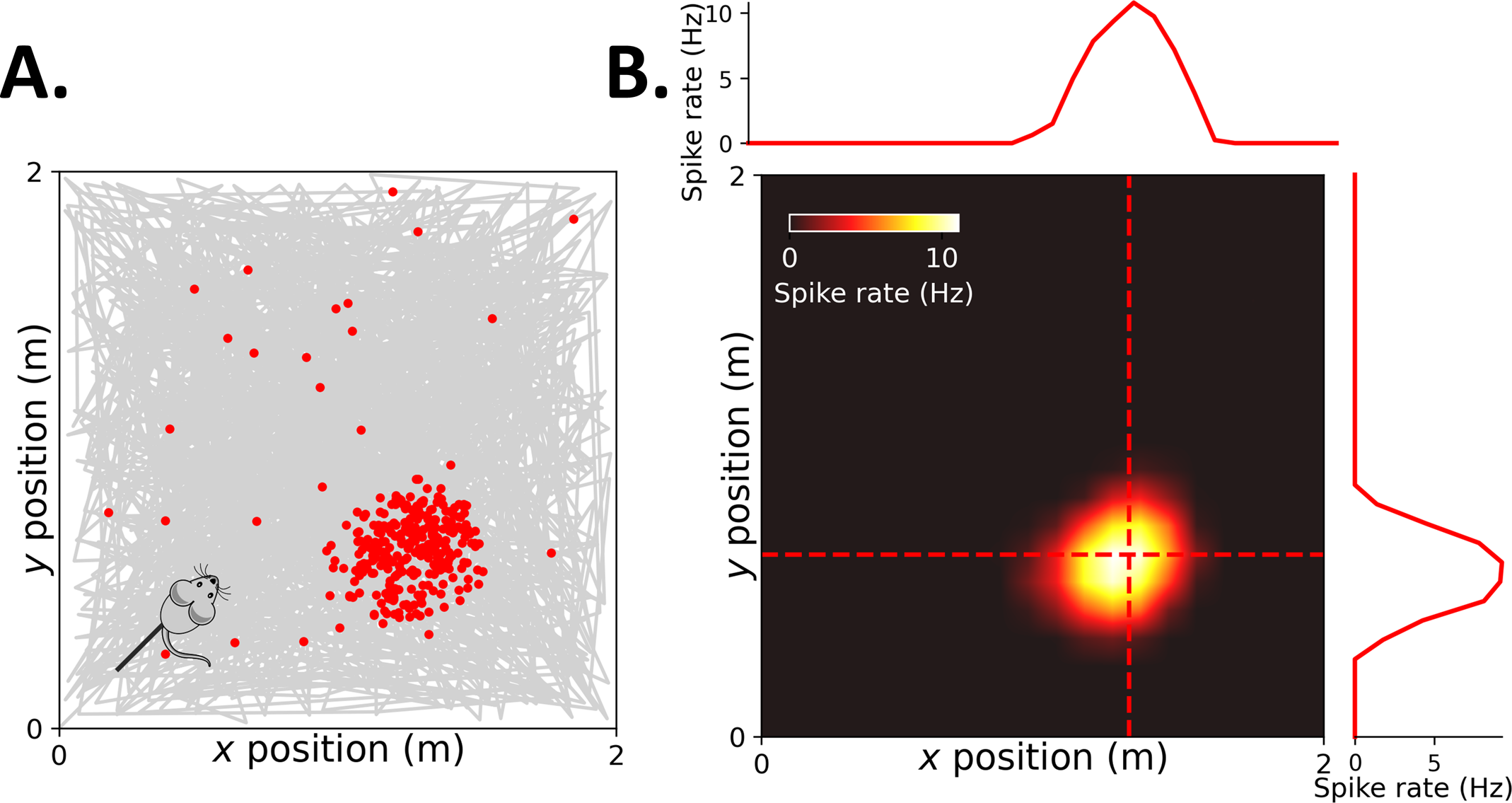}
    \caption{\label{fig:tuningcurves}Simulation showing the trajectory with spiking activity (A)
    together with the spike count map (B), Eq.~\eqref{eq:frmap},
    and its tuning curves (B, top and right).}
\end{figure*}

Neuronal population firing rate responses often depend systematically on specific features of sensory stimuli,
movements, or the animal's internal state. This dependence can be characterized by a \emph{tuning curve},
which describes how the firing rate of a neuron varies with a particular stimulus or behavioral variable.
The value associated with the highest firing rate defines the neuron's \emph{preferred stimulus}.

\paragraph{Spatial coding.}
A classic example is the representation of spatial position by hippocampus neurons.
When a mouse explores an arena, some neurons increase their firing rate whenever the animal enters a particular region.
The region associated with an elevated firing rate is called a \emph{place field}.
By mapping the firing activity onto the animal's position, one obtains a spatial tuning curve,
which describes the firing rate as a function of position, $\rho(x,y)$ (Fig.~\ref{fig:tuningcurves}).
In this case, the firing rate map for a given neuron $i$
can be calculated by adapting Eq.~\eqref{eq:fr_timebinning} to a spatial grid
that discretizes the arena,
\begin{equation}
\label{eq:frmap}
\rho_i(x_k,y_k)=\frac{1}{T(x_k,y_k)}\sum\limits_{t=t_k}^{t_k+T(x_k,y_k)} X_i(t)\ ,
\end{equation}
where $T(x_k,y_k)$ is the time spent on the $(x_k,y_k)$ grid site,
and $t_k$ is the instant in which the mouse enters this site.

Different neurons exhibit different place fields~\cite{OKeefe1971Place},
and these can be remapped according to the environment features,
allowing spatial position to be represented by their firing activity \citep{GoodSmith2017}.

Another type of spatial code is provided by grid cells. These are neurons whose firing rates peak
at multiple locations arranged in a regular, hexagonal grid as an animal explores its environment~\cite{Hafting2005grid}.
This spatial firing pattern provides a representation of Euclidean space and
may support navigation by encoding the animal's position through continuously updated information about movement and direction.

\paragraph{Temporal coding.}
Neuronal activity can also be tuned to elapsed time.
During timing tasks, some neuronal populations exhibit activity patterns associated
with specific intervals or stages of learning. For example,
temporal encoding during the learning of a self-timed response
has been shown to shift from the medial prefrontal cortex
toward the dorsal striatum as animals acquire proficiency in the task \citep{TunesReyes2022TimeCells}.

\paragraph{Visual coding.}
Tuning curves also characterize responses in the visual cortex.
For example, neurons in the primary visual cortex can
respond selectively to the orientation of a moving
bar relative to its direction of motion.
By varying the angle between the bar's orientation and its velocity vector,
one can measure the firing rate as a function of this angle.
The resulting tuning curve identifies the stimulus configuration
that elicits the strongest response \citep{abbottNeuro}.

\paragraph{Motor coding.}
Similar principles apply to motor control. During arm-reaching tasks,
the firing rate of neurons in the primary motor cortex can depend
on the direction in which the animal moves its arm.
By recording neuronal activity across repeated trials and averaging
the firing rate for each reaching direction,
one can construct a tuning curve that reveals the
neurons' preferred movement direction \citep{abbottNeuro}.

\paragraph{Head direction.}
Head-direction cells provide another example of directional tuning.
These neurons respond selectively to the direction in which an animal's head points~\citep{Taube1990HeadDir,Yoder2009HeadDir},
providing information that can contribute to spatial navigation.
Models of navigation can incorporate interactions between head-direction cells
and hippocampal neurons to generate trajectories.

\paragraph{Speed coding.}
Speed cells are neurons whose firing rates correlate with the animal's running speed~\cite{Kropff2015Speed}.
Unlike place cells, their activity is largely independent of visual cues, remaining unaffected by darkness.
Interestingly, their firing rates correlate more strongly with future than with current speed,
suggesting a predictive representation of the animal's movement.

\paragraph{Reward and aversion.}
Neuronal firing rates can also encode the value of stimuli or outcomes.
For example, neurons in reward-related circuits may respond to appetitive stimuli,
such as palatable food, whereas others respond to aversive events.
Such responses provide a rate-based representation of positive and negative experiences \citep{DominguesCarvalho2025}.

\paragraph{Boundary coding.}
Boundary cells are neurons whose firing rates increase when an animal is near an environmental boundary
at a preferred distance and direction.
Initially predicted based on the properties of place cells~\citep{OKeefe1996Border}, these neurons have
since been identified in several regions of the hippocampal formation, including the subiculum, presubiculum, and entorhinal cortex.

Tuning curves describe how physical quantities, such as spatial position, stimulus orientation,
movement direction, reward, elapsed time, speed, etc are encoded into population firing rates.
Thus, they provide a common framework for modeling behavior~\cite{Sargolini2006},
such as spatial navigation, decision making, pattern separation,
from firing rates~\citep{Hopfield1982,HopfieldTank1985Opt1,Gerstner1992AssocMem,Gerstner1997Nav,GerstnerBook2014,Banino2018GridNav,ConeShouval2021,Ghazinouri2024Nav}.

\section{Recurrent feedback and synaptic dynamics in firing-rate models}

The firing-rate description introduced in the previous sections
allows us to examine how recurrent interactions shape population activity over time.
Following \cite{miller2018}, we describe a recurrent excitatory population of neurons
by its mean firing rate $\rho(t)$, an average synaptic activation variable $s(t)$,
and the fraction of available synaptic resources $D(t)$.

We start from our mean-field population dynamics
in continuous time, Eq.~\eqref{eq:GLmodel_conttime}.
We assume that neurons have a fast refractory period, so
the model reduces to Eq.~\eqref{eq:fr_dynamics_neuron} \citep{Ermentrout2010Book}.
The input $I$ is given by the average of Eq.~\eqref{eq:inputrate},
$I(t)=\Iext(t)+\WEE s(t)$,
where $\Iext(t)=\avg{\Iext_i(t)}$ is the external contribution,
and
\begin{equation}
\avg{I^{\rm syn}_i(t)}=\avg{\sum_jG_{ij}\bar{\rho}_j(t)}=\WEE s(t;\rho)
\end{equation}
is an effective synaptic current that depends on the network firing rate $\rho$.
The dimensionless variable $s$ represents
the mean fraction of active synaptic channels ---
a generalization of the model presented in Box~\ref{box:fr_convolution}.
Taking $s(t)=\rho(t)$ with $q=0$ (no inhibitory neurons)
reduces to a pure recurrent excitatory network similar to that in Eq.~\eqref{eq:fr_GLmodelEI},
with very fast refractory period,
\begin{equation}
    \tau\frac{\mathrm{d}\rho}{\mathrm{d}t}
            =-\rho+F\!\left(\WEE s+\Iext\right).
    \label{eq:recurrent_rate}
\end{equation}

Here, we use $F$ as a saturating input--output function of Hill type,
\[
    F(I)=\left\{\begin{array}{ll}
         \rho_0 &\textnormal{ if }I\leq0  \ ,\\
         \rho_0+\rho_{\max}\dfrac{I^{a}}{I^{a}+\sigma^{a}}& \textnormal{ if }I>0\ .
    \end{array}\right.
\]
The parameter $\rho_0\geq0$ specifies the response at zero total
input, and $\rho_{\max}$ is the maximum increase above this value.
Thus, the response approaches $\rho_0+\rho_{\max}$ at large inputs $I$.
The scale $\sigma$ sets the input at which half of this limit
is reached, that is $F(\sigma)=\rho_0+\rho_{\max}/2$.
The exponent $a>1$ controls the steepness of the sigmoid response.
The phenomenological rate equation \eqref{eq:recurrent_rate} describes relaxation
toward $F(I)$, with $\tau$ setting the rate-relaxation time scale.

The synaptic variable $s$ accounts for the effective dynamics
of release and recapture of synaptic resources~\citep{Roth2010Synaptic}.
Synaptic current is activated by presynaptic firing with rate $p_rD\rho$,
where $p_r$ is the release probability of neurotransmitters,
and $D$ accounts for resource availability.
Each release increases activation in proportion to $\alpha_0(1-s)$,
where $\alpha_0$ determines the activation per release event and $1-s$
represents the fraction of channels still available for activation.
Existing activation decays with time constant $\tau_s$, giving
\begin{equation}
    \frac{\mathrm{d}s}{\mathrm{d}t}
    =-\frac{s}{\tau_s}+\alpha_0 p_r D \rho (1-s)\ .
    \label{eq:recurrent_gating}
\end{equation}
The quantity $D$ can be dynamic, in which case activating the synapse
(increasing $s$) would deplete resources, decreasing $D$.
This effect is known as synaptic depression.

\paragraph{Bistability and persistent activity.}
We first examine the recurrent network with fixed synaptic resources $D$.
The system stays stationary under constant external input, so
both the firing rate and synaptic activations are constants, $\rho^*$ and $s^*$,
respectively. These can be determined from Eqs.~\eqref{eq:recurrent_rate}
and~\eqref{eq:recurrent_gating} by making $\mathrm{d}\rho/\mathrm{d}t=\mathrm{d}s/\mathrm{d}t=0$,
resulting in 
\begin{equation}
    \rho^*=F\!\left(\WEE s^*(\rho^*)+\Iext\right)\ ,
    \label{eq:stationary_recurrent}
\end{equation}
where the stationary synaptic current is a function of the firing rate,
\begin{equation}
\label{eq:stationary_syn}
    s^*(\rho)=\frac{\alpha_0 p_r\tau_s D \rho}{1+\alpha_0 p_r\tau_s D\rho}\ .
\end{equation}

\begin{figure*}[t!]
    \centering
    \includegraphics[width=\linewidth]{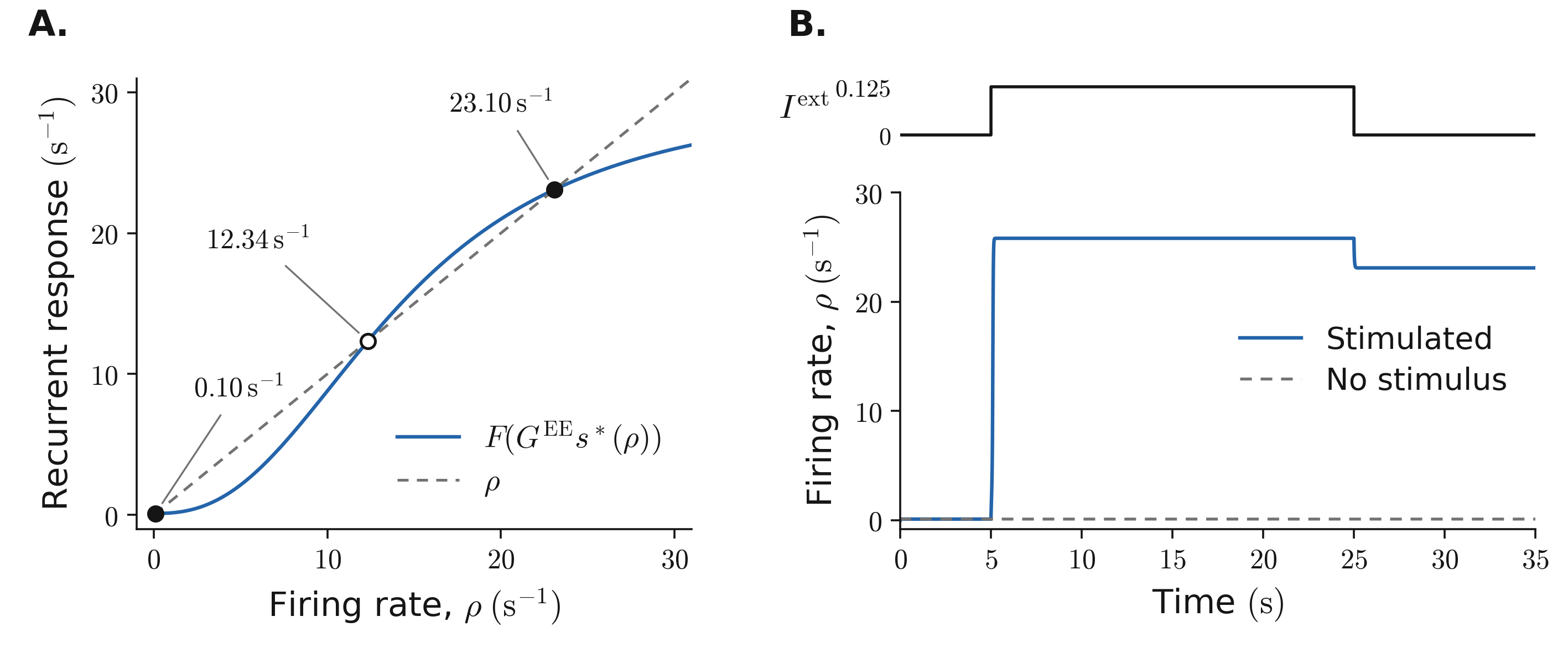}

    \caption{
    Recurrent excitation supports bistability and persistent population activity.
    \textbf{A.} Stationary states of the recurrent network with synaptic resources
    fixed at $D=1$ and $\Iext=0$.
    The curves are both sides of Eq.~\eqref{eq:stationary_recurrent},
    such that their intersection yield the stationary states,
    $\rho^*\approx 0.10$, $12.34$, and $23.10$~s$^{-1}$.
    Filled circles $\to$ stable equilibria;
    open circle $\to$ unstable equilibrium (threshold).
    \textbf{B.} Temporal evolution of the population firing rate.
    The network starts at $\rho=\rho_0$ and after
    receiving an external input rate $\Iext=0.125$
    from $t=5$ to $25\,\mathrm{s}$, it switches
    to the high activity state.
    Without stimulation, the population remains near the lower stable equilibrium $\rho_0$.
    Both simulations start from $\rho(0)=s(0)=0$.
    Fixed parameters: $D=1$,
    $\tau=10\,\mathrm{ms}$,
    $\tau_s=2\,\mathrm{ms}$,
    $\alpha_0=0.5$,
    $p_r=0.2$,
    $\WEE=175$,
    $\rho_0=0.1$~s$^{-1}$,
    $\rho_{\max}=30$~s$^{-1}$,
    $\sigma=0.5$, and
    $a=2.5$.
    }
    \label{fig:recurrent_feedback}
\end{figure*}

We can solve Eq.~\eqref{eq:stationary_recurrent} by looking for
the values of $\rho=\rho^*$ in which the function $f(\rho)=F(\WEE s^*(\rho)+\Iext)$
crosses the identity line $g(\rho)=\rho$ (Fig.~\ref{fig:recurrent_feedback}A).
Setting $\Iext=0$, three intersections
happen\footnote{The intersections also depend on the values
    of $\alpha_0$, $p_r$, $\WEE$, $D$, $\rho_0$, $\rho_{\max}$, $\sigma$ and $a$.
    See the caption of Fig.~\ref{fig:recurrent_feedback} for parameter values.}
at $\rho^*_1=\rho_0=0.1$~s$^{-1}$, $\rho^*_2\approx12.34$~s$^{-1}$, and $\rho^*_3\approx23.1$~s$^{-1}$.
Each of these correspond to a stationary activity state of the network.
Initial conditions lead the network to $\rho^*_1$ or $\rho^*_3$, making them stable.
The intermediary state, $\rho^*_2$, is unstable, since it can only be achieved if we started
the simulation at $\rho(0)=\rho^*_2$, $s(0)=s^*(\rho^*_2)$.

The system exhibits \emph{bistability}, because both
the low- and high-activity states are stable.
In other words, small inputs will only perturb these stationary states.
When inputs are strong enough to make the network firing rate cross
the intermediary (unstable) threshold, the network can switch
from the low to the high activity state, and vice-versa.

For example, Fig.~\ref{fig:recurrent_feedback}B illustrates a network working
at $\rho(t)=\rho^*_1$, and receiving a transient strong stimulus
that crosses the threshold $\rho^*_2$.
When it stops, the network remains active at $\rho(t)=\rho^*_3$
rather than returning to its initial state.
This shows that our simple recurrent network
can sustain either a low- or a high-activity state,
depending on the preceding stimulation. 
This is a simple form of holding a binary memory
about the past inputs~\citep{Khona2022}.

\paragraph{Synaptic depression and oscillatory activity.}

\begin{figure*}[t!]
    \centering
    \includegraphics[width=\linewidth]{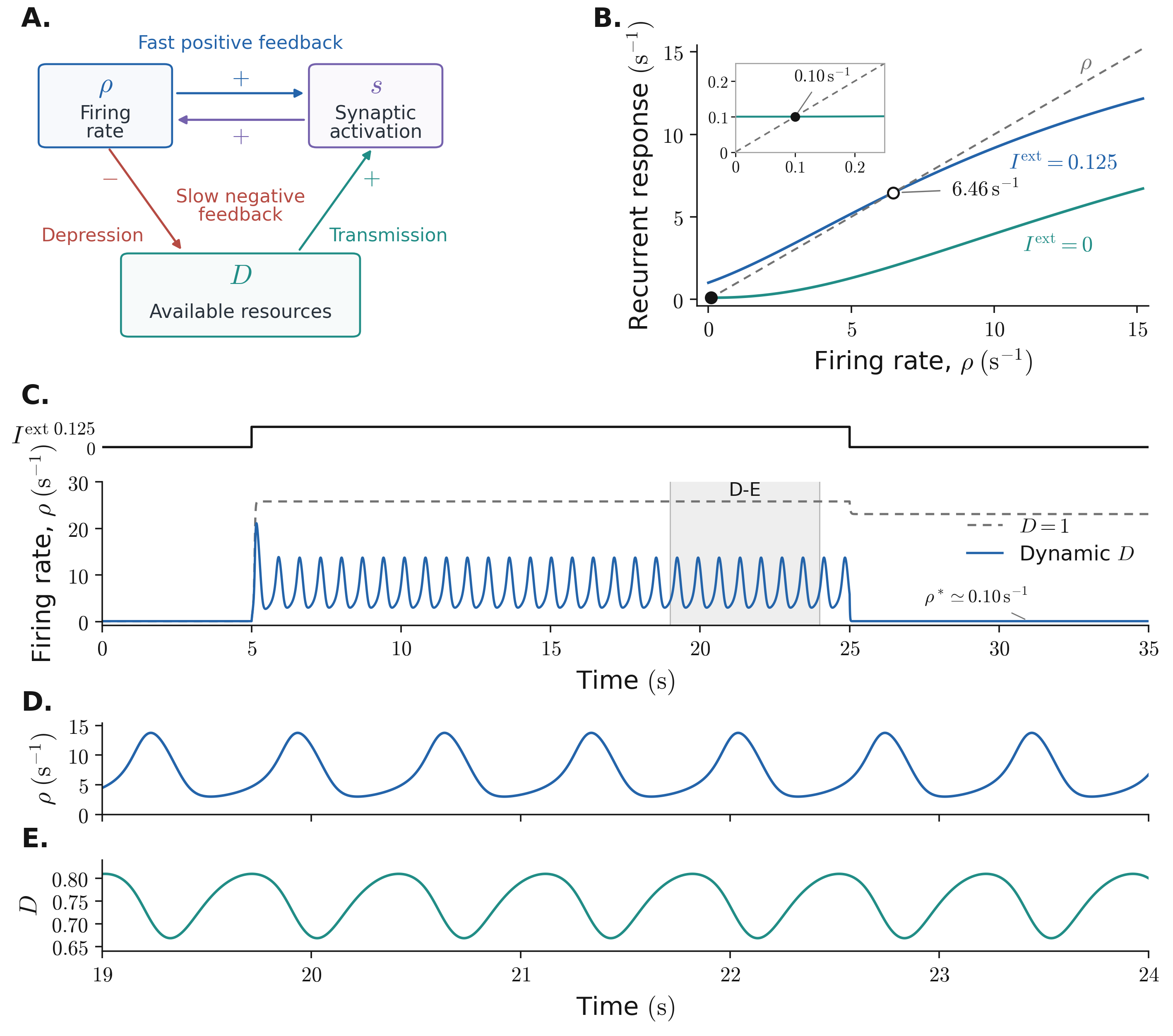}
    \caption{
    Synaptic depression modifies recurrent network dynamics.
    \textbf{A.} Feedback mechanisms: firing rate $\rho$
    promotes synaptic activation $s$ (positive feedback),
    while activity-dependent reductions in available
    resources $D$ weaken recurrent excitation (negative feedback).
    \textbf{B.} Stationary states of the model
    for $\Iext=0$, $\rho^*=\rho_0$ stable, and $\Iext=0.125$,
    $\rho^*\approx6.46$~s$^{-1}$ unstable, are solutions of
    Eq.~\eqref{eq:stationary_recurrent}, with $s^*(\rho)$
    given by Eq.~\eqref{eq:stationary_syn} and $D=D^*(\rho)$
    given by Eq.~\eqref{eq:stationary_resources}.
    \textbf{C.} Comparison of the effect of a constant transient stimulus
    on the population activity $\rho(t)$ for fixed $D=1$ (dashed line)
    and dynamic $D$. The negative feedback of resource
    consumption on synaptic dynamics creates oscillations only
    during stimulation.
    \textbf{D and E.} Detail of the activity $\rho(t)$ and
    $D(t)$ during constant input (shaded region in panel C):
    increasing $\rho$ decreases $D$ and vice-versa, with a period of
    $\approx700$~ms.
    Parameters and initial conditions are as in Fig.~\ref{fig:recurrent_feedback};
    $\tau_D=250\,\mathrm{ms}$.
    }
    \label{fig:synaptic_depression}
\end{figure*}

Let us see what happens when
synaptic resources $D$ are consumed with neuronal firing:
\begin{equation}
    \frac{\mathrm{d}D}{\mathrm{d}t}
    =\frac{1-D}{\tau_D}-p_r D \rho,
    \label{eq:synaptic_resources}
\end{equation}
where $\tau_D$ is the recovery time constant. Presynaptic firing
releases neurotransmitters with probability $p_r$,
so synaptic resources $D$ are consumed with rate $p_r\rho$.
When firing ceases, resources recover toward $D=1$.
A reduction in $D$ weakens synaptic activation $s$, Eq.~\eqref{eq:recurrent_gating},
compromising recurrent excitation (Fig.~\ref{fig:synaptic_depression}A).
Thus, Eq.~\eqref{eq:synaptic_resources}
is a negative feedback into the system.
For our example, we assume that resource recovery is slower
than both the dynamics of firing rate and the synaptic activation,
$\tau_D>\tau>\tau_s$.

Synaptic depression expressed by a dynamic $D$ changes
the stationary state of the network.
The condition $\mathrm{d}D/\mathrm{d}t=0$ gives the
\begin{equation}
\label{eq:stationary_resources}
D^*(\rho)=\frac{1}{1+p_r\tau_D \rho}\ .
\end{equation}
This value goes into Eq.~\eqref{eq:stationary_syn} changing $D\to D^*(\rho)$.
Resource use reduces the stationary synaptic variable $s^*(\rho)$ when compared to the case with
fixed $D=1$, destroying the high-activity stationary states, $\rho^*_2$ and $\rho^*_3$.
Thus, a network with this type of synaptic depression
and no external input, $\Iext=0$, functions at
a low firing rate\footnote{The parameters are the same as those given in Fig.~\ref{fig:recurrent_feedback},
                           with $\tau_D$ given in Fig.~\ref{fig:synaptic_depression}.},
$\rho^*=\rho_0$ (Fig.~\ref{fig:synaptic_depression}B).

During constant stimulation (Fig.~\ref{fig:synaptic_depression}B,C),
the mean firing rate shifts to $\rho^*\approx6.46$~s$^{-1}$.
However, this state is now unstable and an oscillation appears
due to the competition between positive and negative feedback.
Thus, instead of the persistent high-activity state,
the network presents a transient rhythm of
approximately $700\,\mathrm{ms}$ (Fig.~\ref{fig:synaptic_depression}C--E).

During each cycle, increasing $\rho$ promotes synaptic activation $s$
and strengthens recurrent excitation, while reducing the available resources $D$.
But a falling $D$ weakens recurrent excitation and
firing rate declines (Fig.~\ref{fig:synaptic_depression}D--E).
Lower firing rate replenishes the synaptic resources and the cycle restarts.
The oscillation therefore arises from the interaction between recurrent excitation and
resource dynamics under constant input. When the stimulus is removed,
oscillations cease and the population returns to the low-activity
equilibrium (Fig.~\ref{fig:synaptic_depression}C,D).

These examples show that incorporating biologically motivated synaptic
mechanisms can broaden the range of collective behaviors described by
firing-rate models. Recurrent excitation supports
persistent activity with fixed resources. In contrast, resource depletion introduces
a slower negative feedback that supports oscillations under constant input.
Hence, even this simple description helps us understand the role of synaptic
depression on the processing of inputs. Both dynamics studied here
could play different roles in a larger context.
The persistent state network can act as a switch,
whereas the oscillatory one can act as a sort of filter
that allows some inputs to pass, creating rhythmic behavior~\citep{Byrne2020NeuMass}.

\section{What can rates teach us?}
\label{sec:applications}

Here, we present a selection of recent applications
of firing-rate models, illustrating the breadth and diversity of research in this field.

\paragraph{Sequence learning.} \cite{ConeShouval2021} proposed a modular spiking network
of leaky IF neurons that learns both the order and duration of stimulus sequences.
Recurrent connections within each module, particularly those involving Timer cells,
learn stimulus durations, whereas feedforward connections from Messenger cells in one
column to Timer cells in the next encode sequence order.
Learning relies on a local, neuromodulator-gated eligibility-trace rule in which firing rates
determine the Hebbian term. A simplified low-dimensional rate-based model reproduces the main
behaviors qualitatively, linking detailed spiking dynamics to population-level descriptions.

After training, stimulating only the first element is generally sufficient to recall
the learned sequence. Simulations demonstrate the recall of sequences containing up
to eight elements, with reproduced durations typically matching the trained input
within approximately 10\%, despite stochastic variability. The model assigns duration
and order learning to distinct cell populations, Timers and Messengers, whose response
profiles are consistent with V1 recordings. This provides a biologically motivated
alternative to non-local learning methods such as backpropagation through time.

The authors also extend the framework to history-dependent, non-Markovian sequences.
In a rate-based formulation, this extension combines the columnar network with a fixed
recurrent reservoir and a sparse pattern network, while retaining local plasticity.
It offers a computational approach to studying complex sequence learning and illustrates
how rate-based descriptions can capture the computational structure of temporal learning.

\paragraph{Refractory period and criticality.}
\citet{Goychuk2015bicrit} extended the classical Wilson--Cowan
firing-rate model by incorporating stochastic neuronal activity
and \emph{memory}\footnote{Memory refers to non-Markovian stochastic processes,
not the biological function.} effects
associated with refractory periods. They showed
that longer refractory periods can promote population oscillations
and induce transitions to synchronized activity. They also demonstrated
that intrinsic neuronal fluctuations can generate neuronal avalanches
with power-law size distributions, even in large networks. These results
highlight the importance of accounting for neuronal variability and
refractory effects when describing collective brain activity,
particularly oscillations and avalanche dynamics.

\citet{Gong2026Bistab} proposed an extension of the stochastic Wilson--Cowan
model that incorporates memory effects in the refractory periods of excitatory
and inhibitory populations. They showed that prolonged inhibitory refractory
periods can change the nature of the transition to sustained neural activity,
leading to bistability between high- and low-activity states. They also
identified scaling behavior characteristic of directed percolation at the
continuous transition. These findings provide insight into how neuronal
temporal properties influence collective brain states, particularly the
transitions between cortical up and down states observed during sleep and anesthesia.
Criticality and bistability, are believed to be fundamental 
features of cortical networks \citep{Girardi2021aval},
making these works relevant contributions to overall understanding 
of the role of memory in the brain's rate coding and vice-versa.

\paragraph{Preventing seizure-like synchronized states.}
The response of neural networks to internal and external stimuli was investigated
by \cite{Rich2020} using a Wilson--Cowan-based framework. Their model describes
interconnected populations of excitatory--inhibitory (E-I) neurons subject to white noise or periodic inputs.
Despite its relative simplicity compared with more detailed neuronal models,
it reproduces synchronous oscillations and multistability,
which can be interpreted as distinct seizure states \citep{Rhamidda2024SOSync}.

The authors investigated how weak periodic and stochastic inputs modulate the dynamics of E-I populations.
They showed that external stimulation can regulate neural activity, suppress synchronized oscillations,
and prevent the network from entering hyperactive states associated with seizures.

In particular, high-variance and/or high-frequency stimuli,
whether periodic or stochastic, can disrupt multistability
and prevent transitions into seizure states. This regulatory
effect relies on the selective recruitment of inhibitory neurons,
highlighting their role in controlling population activity and maintaining stable neural dynamics.

\paragraph{Jansen--Rit Model.}
The Jansen--Rit model describes the dynamics of a cortical column through interacting
populations of pyramidal neurons and excitatory and inhibitory interneurons.
It is a neural mass model in which neuronal activity is represented by average
firing rates and postsynaptic potentials, coupled through nonlinear transfer functions.
A detailed description of the model can be found in \cite{Knosche2014JREncyclo}
and in the original paper \citep{JansenRit1995}.

As demonstrated by \cite{Spiegler2010}, the Jansen--Rit model exhibits
a rich repertoire of dynamical behaviors, including stationary states,
low-amplitude sinusoidal oscillations arising from Andronov--Hopf
bifurcations\footnote{The Andronov--Hopf bifurcation marks the onset of
        self-sustained oscillations with finite amplitude and well-defined frequency.},
and high-amplitude anharmonic oscillations associated with global bifurcations.
Bistability is also found over broad regions of parameter space, highlighting
the model's ability to describe transitions between distinct patterns of cortical activity.

The model has also been applied to different neuroscientific problems.
\cite{Wendling2000} extended it to reproduce spontaneous epileptiform activity,
providing a framework for investigating the mechanisms underlying seizure
generation and the organization of epileptogenic regions.
\cite{David2006} used the model to interpret event-related
EEG and MEG responses, relating measured signals to the
underlying neuronal populations and pathways involved in sensory processing.

\paragraph{Phase-amplitude coupling in brain rhythms.}
\cite{PowanweLongtin2021} investigate how noise and network
connectivity influence the emergence of oscillations in E-I neural populations.
Studying the Andronov--Hopf bifurcation in a stochastic Wilson--Cowan dynamics,
they show how random fluctuations can generate oscillatory
activity even in the absence of sustained oscillations, and how this activity
changes as the network transitions to a self-sustained oscillatory regime.
Their results provide insight into the mechanisms underlying spontaneous
brain rhythms and the fluctuations in their amplitude,
helping to distinguish noise-driven oscillations from sustained rhythmic activity.
This distinction is relevant for understanding both normal brain rhythms
and abnormal oscillatory activity associated with neurological disorders.

\paragraph{Functionally specific model of inhibition in the visual system.}
\cite{Znamenskiy2024Visual} investigated the organization of inhibition in
balanced network models of the mouse visual cortex. Challenging the classical assumption of diffuse,
uniform inhibition in balanced networks \citep{Vreeswijk1996,Vreeswijk1998} and traditional firing-rate
models such as Wilson--Cowan \citep{WilsonCowan1972}, they showed that parvalbumin-expressing (PV+) interneurons
form stronger inhibitory connections with pyramidal neurons sharing similar visual tuning preferences.
This ``like-to-like'' connectivity provides selective feedback inhibition,
suppressing competing neural activity while stabilizing recurrent network dynamics.

However, functional similarity accounts for only 15--18\%
of the variance in synaptic weights, indicating that anatomical
and other factors also contribute to inhibitory connectivity.
Whether this organizing principle extends to other cortical regions remains an open question.

\paragraph{Large scale brain dynamics.}
Advances in neuroimaging and network science have enabled the development of whole-brain models (WBMs),
which describe the links between brain regions through connectome data \citep{Shimoura2021Conn},
whereas nodes are described by neural mass models or simplified versions of it.
By combining structural connectivity with local neural dynamics,
WBMs reproduce large-scale activity patterns observed in functional
magnetic resonance imaging and electroencephalography.
They provide a framework for investigating brain function in health \citep{haimoviciRestState2013,DallaPorta2026Net}
and disease \citep{Girardi2021Epil,RochaStroke2022},
identifying potential biomarkers, and exploring the effects of therapeutic interventions.

As an example, \cite{Byrne2020NeuMass} developed a WBM based on an extension
of the Wilson--Cowan model to interconnected neural populations.
They used it to investigate the effects of transcranial magnetic
stimulation (TMS) on large-scale brain activity, reproducing
stimulation-dependent changes in functional connectivity,
particularly involving the insula, a region implicated in
altered brain networks associated with depression.

Thus, WBMs provide a means of linking local neural population
dynamics to large-scale brain activity, allowing researchers
to investigate brain function and test the effects of interventions \textit{in silico}.

\section{Are rate-based models real?}
\label{sec:discussion}

What makes a good model of the brain? Models can be organized
along a hierarchy of spatial and temporal scales, from macroscopic descriptions
based on firing rates to microscopic models that explicitly describe membrane
potentials, ionic currents, and the dynamics of individual ion channels. A common
view, particularly in experimental neuroscience, is that more detailed models
are inherently better because they incorporate more of the underlying biology.
However, a model's quality should not be judged solely by its level of detail.
Rather, a good model is one that adequately describes the experimental
observations and answers the question being investigated.

In this sense, the best model of the brain is the one that answers the question we ask.
A model is a tool, and its suitability depends on the phenomenon we want to understand.
It may lack the details necessary to address a particular question, or it may be unnecessarily complex,
making it harder to identify the mechanisms responsible for the observed behavior.
The goal is therefore to develop the simplest model that adequately represents
the phenomenon of interest without sacrificing the features necessary to answer the scientific question.
As \cite{Einstein1934} put it
\begin{quote}
[...] the goal of all theory is to make the irreducible basic elements
as simple and as few as possible without having
to surrender the adequate representation of a single datum of experience.
\end{quote}
Including every biological detail in a model can introduce unnecessary complexity
and obscure the mechanisms underlying the phenomenon we seek to understand.
We should therefore retain only the features relevant to the question at hand,
although identifying those features may itself require extensive trial and error.

The Blue Brain Project provides an illustrative example of the possibilities
and limitations of highly detailed modeling. Initiated in the mid-2000s, the
project aimed to reconstruct and simulate brain tissue at the level of
individual neurons, their morphologies, and their synaptic connections. Such
efforts demonstrate the feasibility of building detailed, biologically grounded
models of neural circuits~\citep{blueBrain2006,Markram2015BlueBrain}. This is
an impressive computational and scientific achievement. However, it also raises
an important question: what can we learn from a model whose complexity
approaches that of the biological system it represents?

Detailed models can serve as valuable tools for testing hypotheses, integrating
experimental data, and exploring the consequences of biological mechanisms.
Nevertheless, their complexity can make it difficult to isolate individual
causal relationships. Understanding a phenomenon requires more than reproducing
its behavior: we also want to determine why that behavior occurs. Ideally, we
should be able to change a parameter, identify the consequences, and relate
those consequences to a specific mechanism.
For example, we may ask how changing
an adaptation time constant or a memory window affects neuronal activity, or
how changing the synaptic coupling strength $J$ alters the coding properties of
a neural network. This is precisely what \cite{kinouchiCopelli}
did when showing that a critical point optimizes network sensibility,
and we reproduced that result here in the mean-field of an E-I network of spiking neurons.
Reduced models can make these relationships explicit and
therefore facilitate mechanistic understanding.

This distinction does not imply that detailed models are unhelpful, nor that
they should replace experimental approaches. Rather, it highlights the
different purposes that models can serve. A detailed simulation may reproduce
the behavior of a neural circuit with remarkable biological realism, while a
simpler model may be better suited to identifying the mechanisms underlying a
particular phenomenon. Neither approach is absolutely superior.

Rate-based models illustrate this principle. If our question concerns how
neurons encode information through their firing rates, we may not need to
describe the ionic currents responsible for generating each action potential.
Instead, we can focus on the relationship between the input to a neuron and its
output firing rate --- the $F$-$I$ curve ---, or on how the output firing rates
of a population depend on the network properties.
Examples include understanding the encoding of an animal's position,
speed, and other features that we described in Sections~\ref{sec:tuningcurve}
and~\ref{sec:applications}.

By contrast, a rate-based model is not appropriate for every problem. If we
want to understand how a particular ionic current generates spike-frequency
adaptation, for example, we need a description of the membrane dynamics that
includes the relevant currents and their associated gating variables. Questions
about the roles of the M-current or the A-current require a model that
explicitly represents their contributions to the membrane potential,
as we have shown in Section~\ref{sec:f-Icurve}. The
appropriate level of description therefore depends on the mechanism we seek to
understand.

This brings us back to the question in the title: are firing rates real? The
fact that firing rates are not fundamental microscopic variables does not make
them arbitrary or physically meaningless. Firing rates are as \emph{real} to
brain function as temperature and pressure are \emph{real} to physical systems:
they are measurable statistical quantities that emerge from the collective
behavior of microscopic units, while also influencing the dynamics across
microscopic and macroscopic scales. Their usefulness depends on whether they
capture the relevant features of the phenomenon under investigation. As
discussed by Gerstner~\citep{Gerstner1992UnivFiringRate} and demonstrated here,
firing rates can provide a unifying description of neural network behavior,
even when the underlying neuronal dynamics differ substantially.

The broader lesson is that biological systems are organized across interacting
spatial and temporal scales. Microscopic processes, such as ion-channel
dynamics, shape neuronal activity; neuronal activity, in turn, gives rise to
population-level dynamics and collective coding properties.
A description at one scale need not reproduce every detail
of the scales below it to be useful, providing
empirically verifiable answers and predictions.

Rate-based models should therefore be understood neither as literal
representations of every aspect of neuronal activity nor as mere mathematical
conveniences devoid of biological meaning. They are effective descriptions
that emerge from the dynamics of neurons and populations, and their validity
must be assessed in relation to the phenomena they are intended to explain.
The important question is not whether firing rates are the only ``real''
description of the brain, but when, why and how they provide answers and predictions
that can be tested.

\section*{Acknowledgments}

M.G.-S.\ acknowledges 
financial support from Fundação de Amparo a Pesquisa e Inovação do Estado de Santa Catarina (FAPESC),
Edital 21/2024 (Grant n.~2024TR002507) and from
Conselho Nacional de Desenvolvimento Científico e Tecnológico (CNPq Grant n.~302102/2025-6).
This research is supported by INCT-NeuroComp (CNPq Grant 408389/2024-9).
This study was financed in part by the Coordenação de Aperfeiçoamento
de Pessoal de Nível Superior -- Brasil (CAPES) -- Finance Code 001
T.T.A.C., D.D.C.~and M.H.~thank partial financial support from CNPq-Brazil,
and C.F.V.~thanks partial financial support from CAPES-Brazil.


\bibliographystyle{abbrvnat}


\end{document}